\documentclass[journal]{IEEEtran}
\ifCLASSINFOpdf
  \usepackage[pdftex]{graphicx}
  \graphicspath{Figures/}
\else
\fi

\usepackage{tikz}
\usepackage{verbatim}
\usetikzlibrary{arrows,shadows,shapes,fit,scopes,calc,matrix,positioning,decorations.pathmorphing}

\usepackage{hyperref}
\usepackage[detect-none]{siunitx}
\usepackage{graphicx}
\usepackage{multirow, array}
\usepackage{subfig, caption, float}
\usepackage{color,soul}
\usepackage{amsmath, mathrsfs, mathtools, amsfonts}
\usepackage{tikz}
\usepackage{xcolor}

\begin{document}
\title{Integrated Approximate Dynamic Programming and Equivalent Consumption Minimization Strategy for Eco-Driving in a Connected and Automated Vehicle}
%
%
%

\author{Shreshta~Rajakumar~Deshpande,
	Daniel~Jung,
	and~Marcello~Canova,~\IEEEmembership{Member,~IEEE}
\thanks{S. Rajakumar Deshpande and M. Canova are with the Center for Automotive Research, Department
of Mechanical and Aerospace Engineering, The Ohio State University, Columbus,
OH, 43212 USA e-mail: (rajakumardeshpande.1@osu.edu, canova.1@osu.edu)}
\thanks{D. Jung is with the Department of Electrical Engineering, Linköping University, Sweden e-mail: (daniel.jung@liu.se).}
}

\markboth{Preprint (Under Review), October~2020}%
{}

\maketitle

\begin{abstract}
This paper focuses on the velocity planning and energy management problems for Connected and Automated Vehicles (CAVs) with hybrid electric powertrains. The eco-driving problem is formulated in the spatial domain as a nonlinear dynamic optimization problem, in which information about the upcoming speed limits and road topography is assumed to be known a priori. To solve this problem, a novel Dynamic Programming (DP) based optimization method is proposed, in which a causal Equivalent Consumption Minimization Strategy (ECMS) is embedded. The underlying vehicle model to predict energy consumption over real-world routes is validated using experimental data. 

Further, a multi-layer hierarchical control architecture is proposed as a pathway to real-time implementation in a vehicle. The DP-ECMS algorithm is introduced for a long-horizon optimization problem, and then adapted for a receding horizon implementation using principles in Approximate Dynamic Programming (ADP). This computationally economical alternative to the traditional DP solution is then benchmarked and evaluated. 
\end{abstract}


%
\IEEEpeerreviewmaketitle

\section{Introduction}
\label{sec::introduction}

Recent years have seen intensive research efforts and rapid advancements in technologies for connected and autonomous driving in the automotive industry. They pledge significant improvements in energy economy, safety and user convenience. In particular, the access to more information, increased computational power, and precision control, have expanded the energy saving potentials of these Connected and Automated Vehicles (CAVs) \cite{ozatay2014cloud, grumert2015analysis}.

Eco-driving typically refers to velocity control for minimizing the energy use over a driving mission \cite{jin2016power, sciarretta2015optimal}. The eco-driving controls use route information such as speed limits, stop sign locations, grade and so on to compute the energy-optimal speed trajectory. Depending on the level of vehicle automation, this velocity trajectory is used as a driver advisory \cite{wan2016optimal} or provided as a reference to the Advanced Driver Assistance Systems (ADAS) such as an Adaptive Cruise Controller (ACC).

When a conventional powertrain (having an internal combustion engine) or a battery electric vehicle (BEV) is considered, the eco-driving control scenario is simpler. A Dynamic Programming (DP) based approach for velocity control is developed for a conventional powertrain in \cite{mensing2013trajectory} and for a BEV in \cite{dib2014optimal}. In \cite{hellstrom2009look,hellstrom2010design}, a real-time velocity optimization algorithm using DP is proposed. The authors in \cite{ozatay2014cloud} solve a DP optimization in the cloud for optimizing the velocity reference that is tracked by a human driver.

Eco-driving applications in vehicles equipped with hybrid electric powertrains present an additional degree of complexity, as an intelligent energy management strategy (EMS) has to be designed along with the velocity optimization. Depending on the approach, the powertrain and velocity controls can be handled separately or optimized simultaneously. In \cite{ozatay2017velocity}, Pontryagin's Minimum Principle is used to solve the optimization problem. In \cite{uebel2017optimal}, a controller that combines the Pontryagin's Minimum Principle with DP is proposed. The look-ahead optimization scheme in \cite{olin2019reducing} utilizes DP for combined velocity and torque split optimization in a mild-hybrid electric vehicle. The DP-based approach in \cite{heppeler2014fuel} considers intelligent battery management along with velocity optimization. A Model Predictive Control (MPC) based scheme is developed in \cite{sun2014velocity}, that utilizes a neural network to predict the velocity profile. In \cite{xiang2017energy} as well, neural networks are used for predicting the velocity, with the energy management handled by a nonlinear MPC that uses a forward dynamic programming algorithm. An emerging area of interest is the use of reinforcement learning for eco-driving control, as in \cite{liu2017reinforcement}.

In literature, a hierarchical decision system is popularly adopted in an eco-driving control scenario \cite{homchaudhuri2016fast, sun2018robust, lim2016distance}. In \cite{heppeler2017predictive}, a multi-layer scheme is considered for optimizing the velocity and energy management over long and short distance ranges, with an on-line Equivalent Consumption Minimization Strategy for real-time powertrain control. This type of architecture is suitable for autonomous systems to handle the complex control problem of combining vehicle motion planning, velocity trajectory optimization and powertrain control \cite{paden2016survey}. When a travel route is selected, the velocity can be optimized with respect to route information, traffic, and the powertrain energy management system, such that the predicted fuel consumption is minimized. The optimized velocity trajectory, together with the resulting energy management strategy, are then used as input to the low-level real-time control of the vehicle.


\section{Objective and Problem Formulation}
\label{sec::objective_problem}

The objective in this work is to develop a control system for autonomous velocity optimization and powertrain control of a mild-hybrid electric vehicle (mHEV). It is assumed that GPS data and route information, such as speed limits, road slope, and locations of stop signs, are available. Further, surrounding traffic is implicitly handled by a traffic management system that is assumed to control the traffic flow with variable speed limits to avoid collisions \cite{grumert2017using}. For adapting to such varying route conditions, it is necessary to continuously update the velocity optimization and energy management strategy during the driving mission.

To this end, a nonlinear dynamic optimization is formulated to minimize the fuel consumption of the vehicle over the entire driving mission by controlling its velocity trajectory and torque split strategy. This problem is cast in the spatial domain to simplify the process of incorporating position-based route information. Consider a discrete-time dynamic control problem having the form:
\begin{equation}
x_{k+1} = f_k(x_k,u_k) \quad \forall k = 1, \ldots, N-1
\end{equation}
where $k$ is the grid point along the route, $x_k \in \mathcal{X}_k \subset \mathbb{R}^n$ is the state, $u_k \in \mathcal{U}_k \subset \mathbb{R}^m$ is the control input, and $f_k$ is the state transition function. Typical control inputs in optimal eco-driving and classical HEV energy management problems include the engine and electric motor torques \cite{olin2019reducing, sciarretta2015optimal}. The control action and state are constrained by a function $h_k: \mathcal{X}_k \times \mathcal{U}_k \to \mathbb{R}^r$ that takes the form:
\begin{equation}
h_k(x_k,u_k) \leq 0, \quad \forall k = 1, \dots, N-1
\end{equation}
This may include route speed limits, operating limits of physical actuators and subsystems, constraints for drive comfort and so on. An admissible control map at grid point $k$ is defined as a map $\mu_k : \mathcal{X} \to \mathcal{U}$ such that $h(x,\mu_k(x)) \leq 0, \forall x \in \mathcal{X}$. The collection of admissible control maps, $\mathcal{M} := \left(\mu_{1}, \dots, \mu_{N-1} \right)$ is referred to as the control policy. The controller aims at minimizing a cost, given by:
\begin{equation}
\label{eq::prb_cost_fn_gen}
\begin{aligned}
J(\mathcal{M}) = g_N(x_N) + \sum_{k = 1}^{N-1} g_k(x_k,u_k)
\end{aligned}
\end{equation}
where $g_k : \mathcal{X}_k \times \mathcal{U}_k \to \mathbb{R}$ is the per stage cost function, considered in this work as a weighted average of the fuel consumption and travel time:
\begin{equation}
\label{eq::prb_cost_fn}
\begin{aligned}
g_k(x_k,u_k) = \left(\gamma \cdot \frac{\dot{m}_{f,k}(x_k,u_k)}{\dot{m}_f^{norm}} + (1-\gamma)\right) \cdot t_k \\
\end{aligned}
\end{equation}
where $\gamma$ is a tuning parameter that represents driver aggressiveness by penalizing the travel time, $\dot{m}_{f,k}$ is the fuel consumption rate, $\dot{m}_f^{norm}$ is a cost normalizing factor and $t_k$ is the per stage travel time.

\section{Motivation and Proposed Contributions}
\label{sec::motivation_contrib}

\begin{figure*}[!t]
	\centering
	\includegraphics[width=1.5\columnwidth]{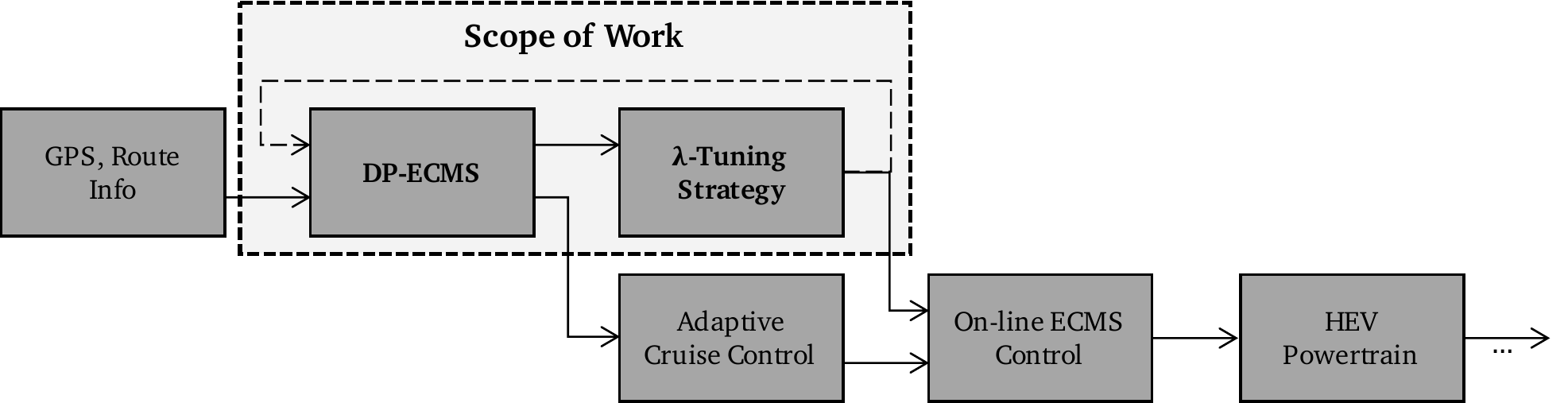}
	\caption{Simplified rendition of proposed control architecture for eco-driving applications.}
	\label{fig::DPECMS_ctrl_architecture}
\end{figure*}



Dynamic Programming (DP) has been used extensively in literature to solve numerically HEV energy management and eco-driving problems, due to its ability to handle discontinuities, integer states and nonlinearities in states or control. DP uses the Bellman equation to break down an optimization problem into a sequence of sub-problems, and solve them in a backward manner \cite{bellman1966dynamic}. Starting from the terminal stage, whose cost is defined as $J_N(x_N) = g_N(x_N)$, the intermediate calculation steps in DP are given by the following recursion \cite{guzzella2007vehicle}:
\begin{equation}
\label{eq::bellman}
\begin{aligned}
J_k(x_k) = \min_{\mu_k(x_k)} \quad J_{k+1}(f_k(x_k,\mu_k(x_k))) + g_k(x_k,\mu_k(x_k)), & \\
\forall k = 1,\dots,N-1 &
\end{aligned}
\end{equation}
where the optimal control policy $\mathcal{M}^* = \left(\mu_{1}^*, \dots, \mu_{N-1}^* \right)$ for an initial condition $x_1$ is given by the trajectory that minimizes $J_1(x_1)$, and $J_{k+1}(f_k(x_k,u_k))$ is the optimal cost-to-go from the projected state to the terminal state. In this way, the cost functional defined in \eqref{eq::prb_cost_fn_gen} is minimized systematically using a constrained DP algorithm. A general drawback of DP is the computational cost that grows exponentially with the dimensions of the state and action spaces, a phenomenon Bellman refers to as the curse of dimensionality \cite{bellman1964dynamic}. The total number of computations in DP is given by:
\begin{equation}
\label{eq::computation_DP}
n_{c} = \mathcal{O} \left(N \cdot \prod_{i = 1}^{n} n_{x,i}  \cdot \prod_{i = 1}^{m} n_{u,i} \right)
\end{equation}
where $N$ is the number of stages, $n_x$ and $n_u$ are the number of discretized points in the state and control grids respectively, and $n,m$ refers to the dimensions of the state and action spaces respectively \cite{Larson:1982:PDP:578147}. To reduce the computational cost, models of HEVs for fuel economy optimization are often limited to a few states (vehicle velocity, battery State-of-Charge, and selected gear, for example).

An alternative to DP for on-line hybrid powertrain energy management is the Equivalent Consumption Minimization Strategy (ECMS). The ECMS is a real-time heuristic control strategy derived from the Pontryagin's Minimum Principle (PMP) by assuming that the co-state in the PMP formulation remains constant over the trip \cite{paganelli2002equivalent}. In the ECMS, the objective function comprises of the actual fuel consumption and an equivalent cost term that represents the electrical energy consumption from the battery. The following instantaneous minimization problem is solved at each stage $k$:
\begin{equation}
\label{eq::ECMS_opt_gen}
\begin{aligned}
\min_{\bar{u}_k} \quad \dot{m}_{f,k}(x_k,\bar{u}_k) + \lambda \cdot f_{pen,k}(x_k) \cdot \frac{P_{batt,k}^{des}(x_k,\bar{u}_k)}{Q_{lhv}}, &\\
\forall k = 1,\dots,N-1 &
\end{aligned}
\end{equation}
where $\bar{u}_k$ is the control input optimized by the ECMS (essentially the torque split), $Q_{lhv}$ is the energy density of the fuel (lower heating value), ${P_{batt,k}^{des}}$ is the power requested from the battery, $\lambda$ is the equivalence factor, a trade-off parameter that defines the battery-fuel equivalence and represents the future fuel cost needed for recharging the battery, and $f_{pen,k}$ is a SoC-dependent penalty function to ensure boundedness of the SoC state. The equivalence factor is route-dependent and needs to be tuned for achieving charge sustaining behavior. Adaptive tuning strategies have been proposed, see for instance \cite{musardo2005ecms, pisu2007comparative}.

\subsection{Proposed Contribution}
In this work, a novel control architecture is proposed for application to eco-driving in a HEV, as shown in Fig. \ref{fig::DPECMS_ctrl_architecture}. This  architecture harnesses and further extends the benefits from both the DP and ECMS optimization schemes while reducing the computational and calibration effort required for real-time implementation. The DP-ECMS approach allows one to reduce the dimensions of the state and action spaces, thereby significantly reducing the computational effort with minimal loss in optimality.


One of the outputs from the DP-ECMS algorithm is the optimal vehicle velocity that can be fed as a reference for an on-board Adaptive Cruise Controller (ACC). Furthermore, the DP-ECMS algorithm and the $\lambda$-tuning strategy determine the optimal equivalence factor. This is used by the on-line ECMS controller based on \eqref{eq::ECMS_opt_gen}, along with the desired powertrain torque request from the ACC, to optimally determine the HEV powertrain torque split strategy. This implementation can significantly reduce the calibration effort to integrate the eco-driving controls with the existing on-board vehicle controllers.

\subsection{Reformulation of Optimization Problem}
\label{sec::problem_reformulation}

The problem discussed in Section \ref{sec::objective_problem} is now reformulated with additional details. The state variables used in the DP-ECMS optimization are the energy (or more precisely, the square of the vehicle velocity, $v_{k}^2$) and the battery SoC ($\xi_k$). One of the key benefits of using the energy as the state instead of the velocity is that the simple Euler forward method can be used with adequate solution characteristics, as discussed in \cite{hellstrom2010design}.

The typical control variables of choice when DP is employed to solve the classical HEV energy management problem are the engine brake torque ($T_{eng,k}$) and BSG torque ($T_{bsg,k}$). Here, at each discretized (state) grid point the DP explores several combinations of these two discretized control inputs. A computationally economical alternative to this approach is adopted in the proposed DP-ECMS algorithm, where this exploration is performed over the discretized powertrain torque instead, while the torque split is optimized by the ECMS. A clear benefit of this structure is that the overall optimization problem is still cast within the DP framework (which returns robust, closed-loop optimal policies), while reducing the dimension of the action space. The optimization states and control actions are summarized as follows:
\begin{equation}
\begin{aligned}
x_k = \begin{bmatrix}
v_{k}^2 \\
\xi_k
\end{bmatrix} \in \mathcal{X}_k \subset \mathbb{R}^2,
\hspace{0.5cm}
u_k = \begin{bmatrix}
T_{pt,k}
\end{bmatrix} \in \mathcal{U}_k \subset \mathbb{R}
\label{eq::states_inputs_DP_ECMS}
\end{aligned}
\end{equation}
where $T_{pt,k}$ is the powertrain torque. In the ECMS developed in this work, the control variable considered is the electric motor torque (specifically, the Belted Starter Generator or BSG torque): $\bar{u}_k = [T_{bsg,k}] \in \bar{\mathcal{U}}_k \subset \mathbb{R}$. Further, the cost index from \eqref{eq::prb_cost_fn_gen} that the DP-ECMS controller aims to minimize is reformulated as shown below:
\begin{equation}
\label{eq::prb_cost_fn_gen_DPECMS}
\begin{aligned}
J(\mathcal{M}; \lambda) = g_N(x_N) + \sum_{k = 1}^{N-1} g_k(x_k,u_k ; \lambda)
\end{aligned}
\end{equation}
where $g_k$ is the per stage cost function that is now parametrized by the equivalence factor $\lambda$ as the DP framework now contains the ECMS algorithm as well.



\section{Model of Parallel Mild-Hybrid Vehicle}
\label{sec::model}

\begin{figure}[!htbp]
	\centering
	\includegraphics[width=\columnwidth]{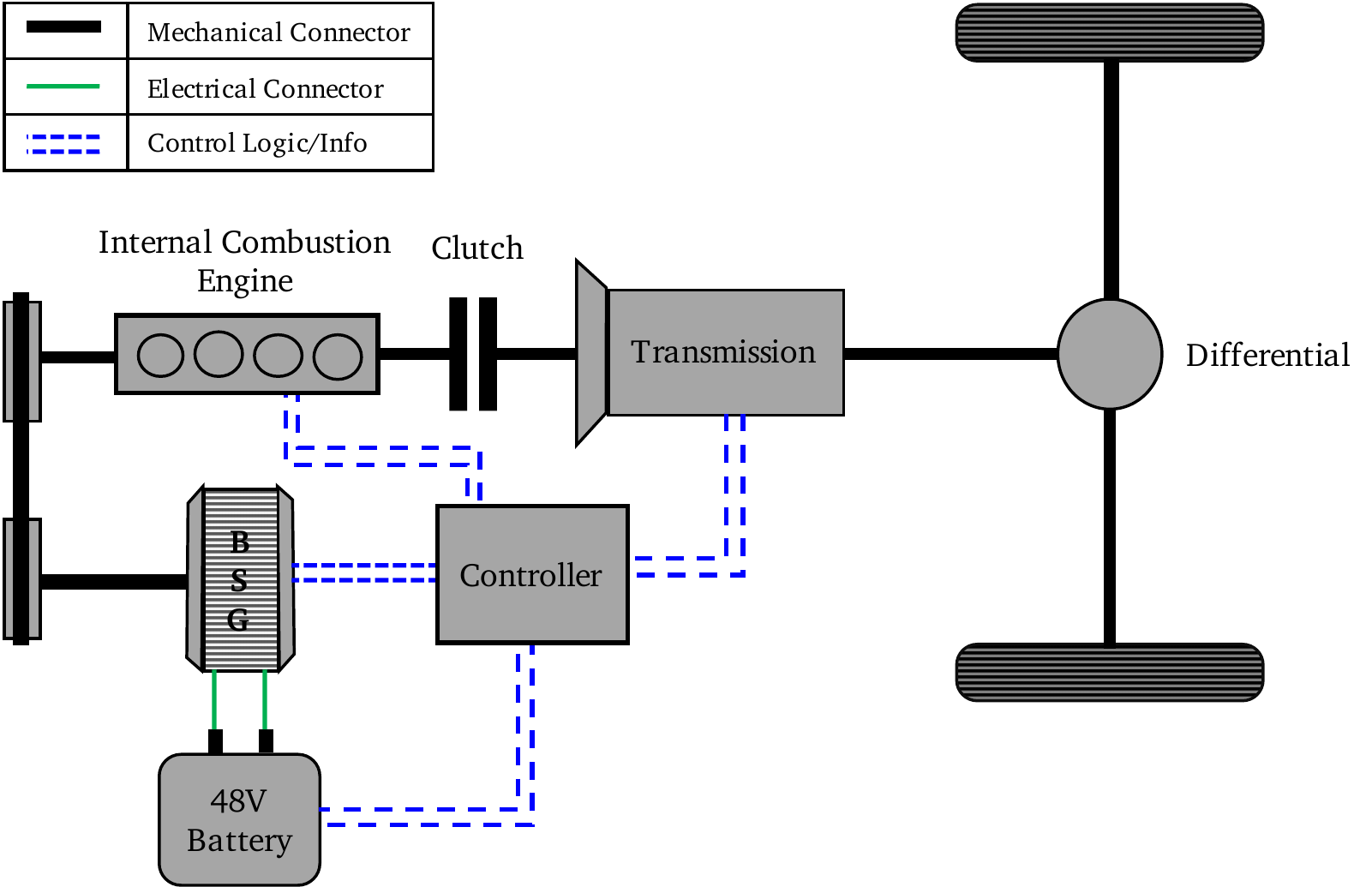}
	\caption{Block diagram of P0 mHEV topology.}
	\label{fig::NEXTCAR_mHEV_topology}
\end{figure}

The parallel mHEV topology studied in this work is illustrated in Fig. \ref{fig::NEXTCAR_mHEV_topology}. A Belted Starter Generator (BSG) replaces the conventional alternator, and is connected to the crankshaft of a 1.8L turbocharged gasoline engine and a 48V battery pack. 

\begin{figure}[!htbp]
	\centering
	\includegraphics[width=\columnwidth]{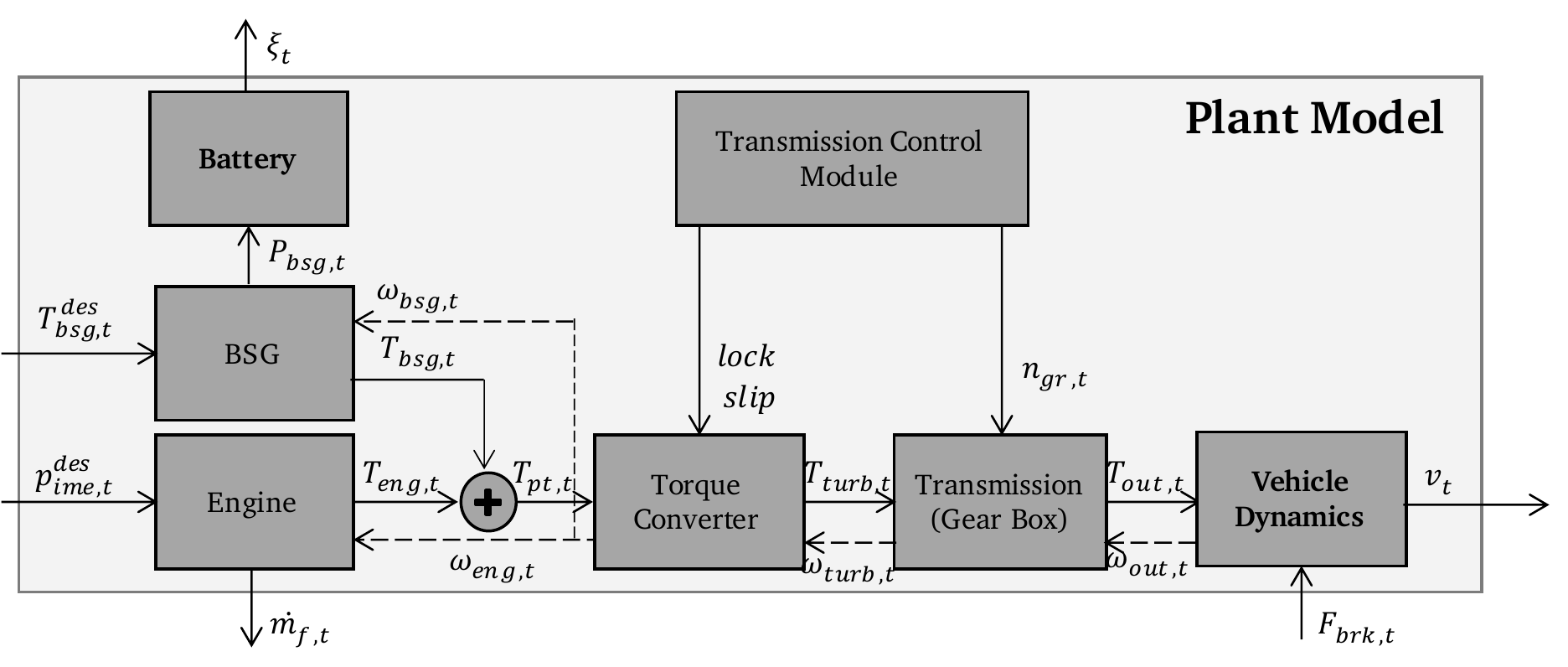}
	\caption{Block diagram of 48V P0 mild-hybrid drivetrain.}
	\label{fig::plant_model}
\end{figure}

A forward-looking dynamic powertrain model is developed for fuel economy evaluation and control strategy verification over prescribed routes. It contains the following key elements:
\begin{itemize}
	\item Quasi-static models of the engine and BSG;
	\item Low-frequency, dynamic model of the battery;
	\item Quasi-static models of the torque converter and transmission;
	\item Low-frequency model of the vehicle longitudinal dynamics.
\end{itemize}
The structure of this forward model has been depicted in Fig. \ref{fig::plant_model}. The inputs to the plant model are the desired BSG torque ($T_{bsg,t}^{des}$) and desired IMEP ($p_{ime,t}^{des}$). These are obtained from a (simplified) model of the Electronic Control Unit (ECU), that contains a baseline torque split strategy and other essential functions that allow conversion from driver’s input (pedal position) to powertrain commands.

\subsection{Engine Model}
The engine fuel consumption is modeled as a static nonlinear map $\psi_t(\cdot,\cdot)$ of engine torque and speed. The brake mean effective pressure, actual brake torque and fuel flow rate obtained from this model can be respectively expressed as:
\begin{equation}
\label{eq::engine_model_T_eng_fuel}
\begin{aligned}
p_{bme,t} &= p_{ime,t}^{des} - p_{fme,t}(T_{eng,t}^{des},\omega_{eng,t}) - p_{pme,t}(\omega_{eng,t})\\
T_{eng,t} &= p_{bme,t} \cdot \frac{V_d}{4\pi} \\
\dot{m}_{f,t} &= \psi_t(T_{eng,t},\omega_{eng,t})
\end{aligned}
\end{equation}
where, $\omega_{eng,t}$ is the engine speed, $p_{fme,t}$ is the friction mean effective pressure and $p_{pme,t}$ is the pumping mean effective pressure. These maps are based on steady-state engine test bench data provided by a supplier; the same maps and torque structure are used in the simplified ECU model.

\subsection{BSG Model}
A simplified, quasi-static model is used to compute the BSG torque and electrical power output:
\begin{equation}
\label{eq::BSG_model}
\begin{aligned}
\omega_{bsg,t} &= r_{belt}\cdot \omega_{eng,t}\\
P_{bsg,t} &= T_{bsg,t} \cdot \omega_{bsg,t} \cdot \bar{\eta}_{bsg,t} \\
\bar{\eta}_{bsg,t} &= {\begin{cases}
	\eta_{bsg,t}(\omega_{bsg,t},T_{bsg,t}), & T_{bsg,t} < 0 \\
	\frac{1}{\eta_{bsg,t}(\omega_{bsg,t},T_{bsg,t}) }, & T_{bsg,t} > 0
	\end{cases}}
\end{aligned}
\end{equation}
where $r_{belt}$ is the belt ratio, $P_{bsg,t}$ is the electrical power required to produce a torque $T_{bsg,t}$ at speed $\omega_{bsg,t}$, and $\eta_{bsg,t}$ is the map-based BSG efficiency.

\subsection{Battery Model}
A zero-th order equivalent circuit model is used to compute the battery State-of-Charge (SoC) and voltage. The model equations are:
\begin{equation}
\label{eq::battery_model}
\begin{aligned}
I_{batt,t} &= \frac{V_{oc,t}(\xi_t) - \sqrt{V^2_{oc,t}(\xi_t) -4R_0\cdot P_{bsg,t}}}{2R_0} \\
\bar{I}_{batt,t} &= I_{batt,t} + I_{bias} \\
\frac{\mathrm{d}\xi_t}{\mathrm{d}t} &= -\frac{1}{C_{nom}}\cdot \bar{I}_{batt,t}
\end{aligned}
\end{equation}
where $V_{oc,t}$ is the battery open-circuit voltage, $R_0$ is a map-based approximation of the battery internal resistance, $I_{batt,t}$ is the battery current, $\xi_t$ is the battery SoC, and $C_{nom}$ is the nominal capacity of the battery. Further, a calibration term $I_{bias}$ is introduced as a highly simplified representation of the on-board electrical auxiliary loads.

\subsection{Torque Converter Model}
A simplified torque converter model is developed with the purpose of computing power losses during traction and regeneration modes. The model equations are:
\begin{equation}
\label{eq::TC_model}
\begin{aligned}
\omega_{p,t} &= \omega_{turb,t} + \omega_{slip,t}^{des}(n_{gr,t}(v_t,T_{eng,t}),\omega_{eng,t},T_{eng,t}) \\
\omega_{eng,t} &= \begin{cases}
\omega_{p,t}, & \omega_{p,t} \geq \omega_{eng,stall} \\
\omega_{idle}, & 0\leq \omega_{p,t} < \omega_{eng,stall} \\
0, & 0\leq \omega_{p,t} < \omega_{eng,stall} , stop = 1
\end{cases}\\
T_{turb,t} &= T_{pt,t}
\end{aligned}
\end{equation}
where $\omega_{p,t}$ is the speed of the torque converter pump, $\omega_{turb,t}$ is the speed of the turbine and $\omega^{des}_{slip,t}$ is the torque converter clutch slip (determined by the ECU based on engine operating condition); $\omega_{eng,stall}$ is the speed at which the engine stalls, $\omega_{idle}$ is the idle speed (target) of the engine, $stop$ is a flag from the ECU indicating engine shut-off when the vehicle is stationary, $T_{turb,t}$ is the turbine torque, and $T_{pt,t}$ is the powertrain torque.

\subsection{Transmission Model}
The transmission is modeled as a static gearbox with efficiency $\eta_{trans,t}$, which is determined empirically from vehicle test data:
\begin{equation}
\label{eq::transmission_model}
\begin{aligned}
\omega_{turb,t} &= r_{f} \cdot r_{gr,t}(n_{gr,t}(v_{t},T_{eng,t})) \cdot \frac{v_{t}}{R_w} \\
T_{out,t} &= r_{f} \cdot r_{gr,t}(n_{gr,t}(v_{t},T_{eng,t})) \cdot T_{turb,t} \cdot \bar{\eta}_{tran,t}\\
\bar{\eta}_{tran,t} &= \begin{cases}
\eta_{tran,t}(\omega_{turb,t},T_{turb,t}), & T_{turb,t} \geq 0 \\
\dfrac{1}{\eta_{tran,t}(\omega_{turb,t},T_{turb,t})}, & T_{turb,t} < 0
\end{cases}
\end{aligned}
\end{equation}
where $r_{f}$ is the final drive ratio, $r_{gr,t}$ is the gear ratio, $R_w$ is the rolling radius of the wheel, and $T_{out,t}$ is the transmission output shaft torque.

\subsection{Vehicle Longitudinal Dynamics Model}
This model is based on the road-load equation, in which only longitudinal dynamics is considered to obtain the balance of tractive force at the wheel and road-load \cite{guzzella2007vehicle}:
\begin{equation}
\label{eq::longitudinal_dyn_model}
\begin{aligned}
\frac{\mathrm{d}v_t}{\mathrm{d}t} &= \frac{F_{trc,t} - F_{road,t}(v_t)}{M}\\
F_{trc,t} &= \frac{T_{out,t}}{R_w} - F_{brk,t}
\end{aligned}
\end{equation}
where $v_t$ is the velocity of the vehicle, $M$ is the mass of the vehicle, $F_{trc,t}$ is the net force exerted by the propulsion system (including braking force, $F_{brk}$) of the vehicle and $F_{road,t}$ is the road-load; it is defined as the force imparted on a vehicle while driving at constant speed over a smooth level surface from sources such as tire rolling resistance, driveline losses, and aerodynamic drag:
\begin{equation}
\label{eq::road_load}
\begin{aligned}
F_{road,t}\left(v_{t} \right) =  \dfrac{1}{2}C_x\rho_a A_f v_{t}^2 &+ Mg \cos{\alpha} \cdot C_{r,t}\left(v_{t} \right) \\
& + Mg\sin{\alpha}
\end{aligned}
\end{equation}
where $C_x$ is the aerodynamic drag coefficient, $\rho_a$ is the air density, $A_f$ is the effective aerodynamic frontal area, $C_r$ is rolling resistance coefficient, and $\alpha$ is the road grade (expressed in \emph{rad}).

\subsection{Model Verification}

\begin{figure}[!htbp]
	\centering
	\vspace{-3mm}
	\subfloat[Vehicle speed and battery SoC comparison]{\includegraphics[width=\columnwidth]{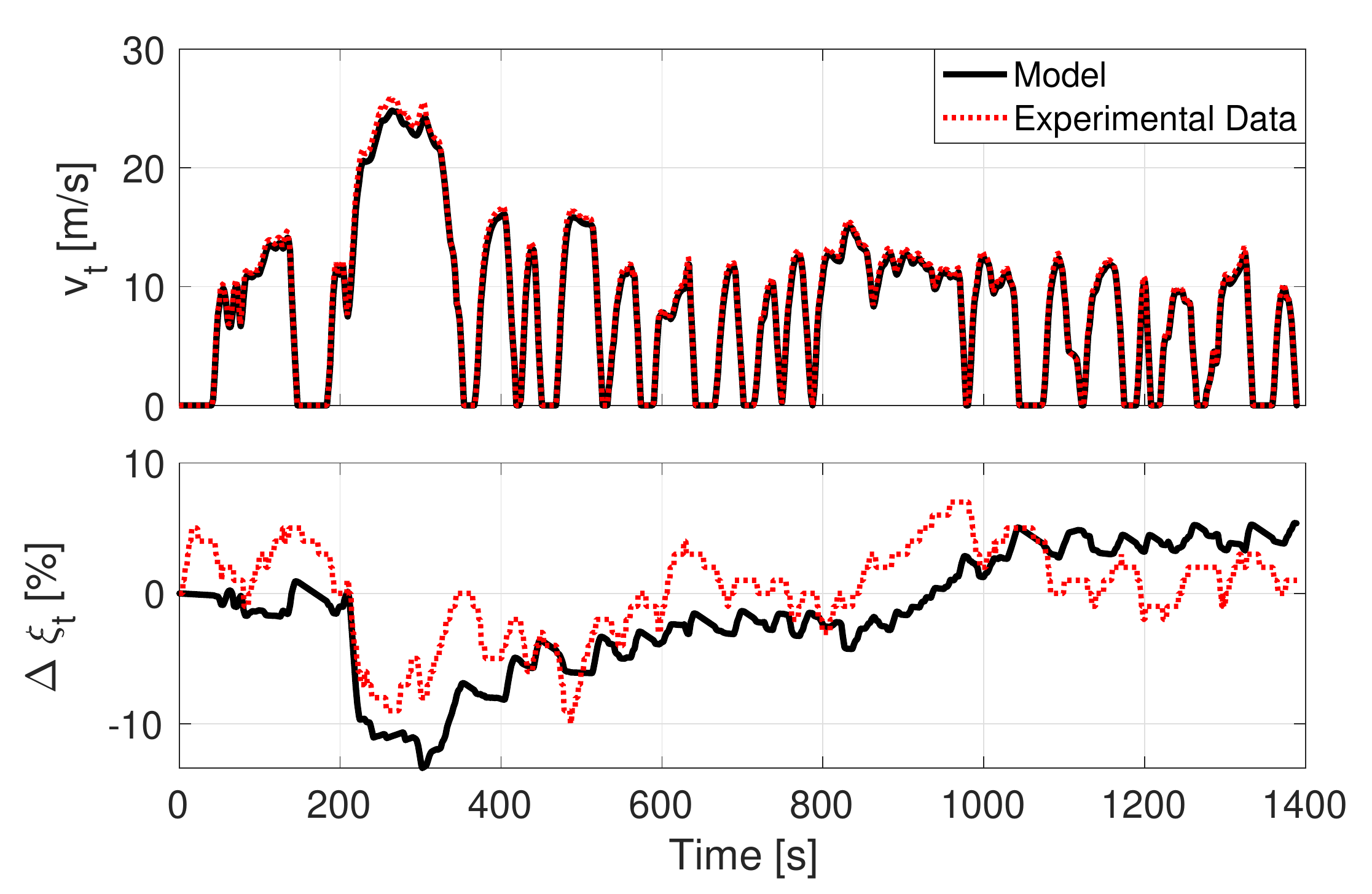}}
	\hfil
	\subfloat[Cumulative fuel consumption comparison]{\includegraphics[width=0.7\columnwidth]{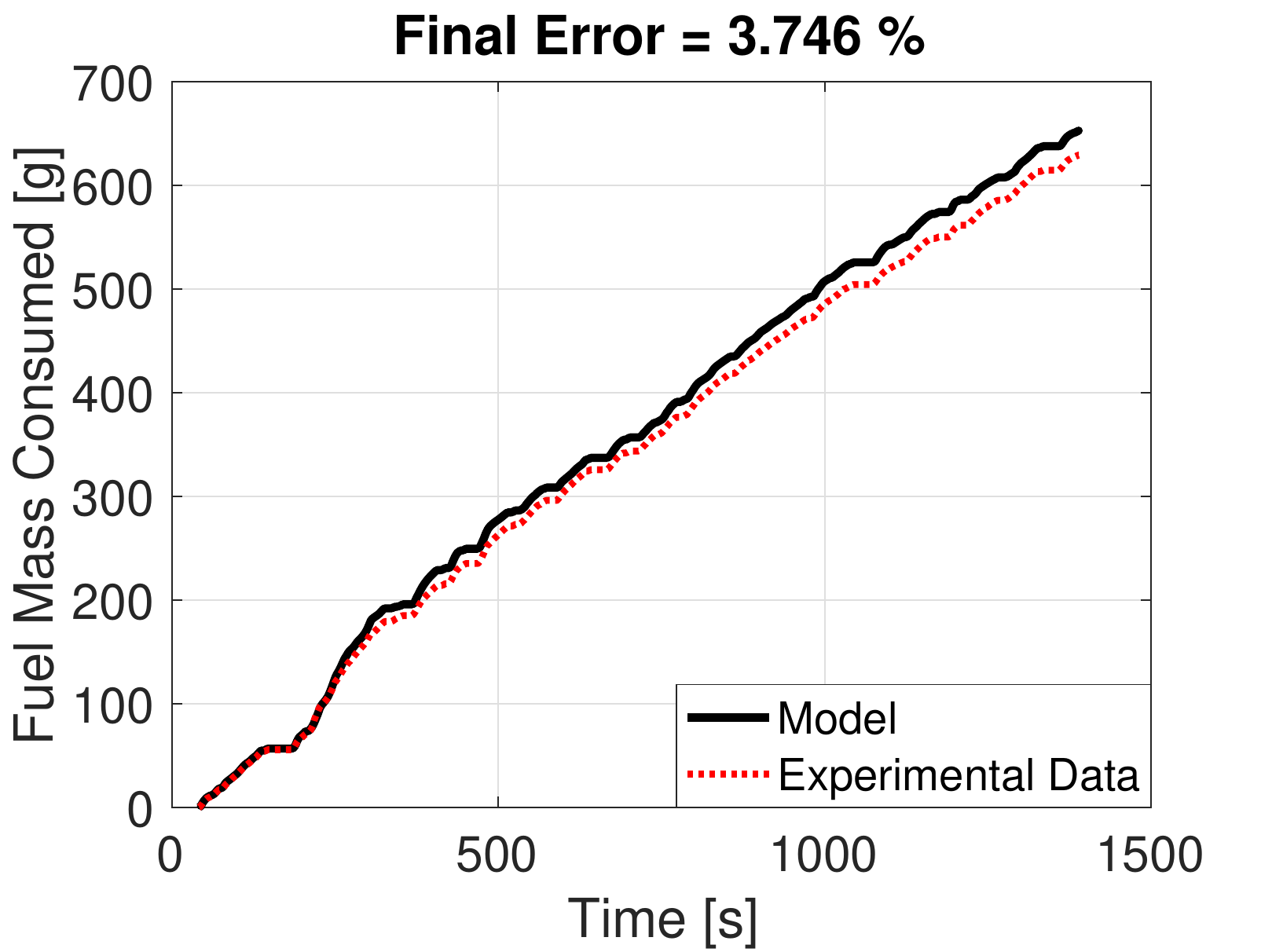}}
	\caption{Validation of forward vehicle model over FTP cycle.}
	\label{fig::model validation_FTP_results}
	\vspace{-1mm}
\end{figure}

The forward model is calibrated and verified using experimental data for the FTP regulatory drive cycle. Fig. \ref{fig::model validation_FTP_results} shows the results of vehicle model verification over the FTP regulatory drive cycle. The experimental data for validation is obtained from chassis dynamometer testing.


The fuel consumed over the cycle is well estimated by the model, with error on the final value less than 4\% relative to the actual engine. Mismatches in fuel economy are caused by differences in the model of the production powertrain control strategy, which does not account for the different calibrations of engine and BSG during cold start conditions.  In lieu of the approximations made, the calibration is considered satisfactory for the purpose of predicting fuel consumption and battery SoC profiles over user-defined routes.
\section{Full-Route Optimization using DP-ECMS}
\label{sec::DPECMS_const_lambda_FR}

To perform the constrained optimization problem described in Section \ref{sec::motivation_contrib}, a custom DP-ECMS algorithm is developed and employed. The reader should note that the notations defined in Section \ref{sec::objective_problem} are used here as well, with frequent references to the DP and ECMS equations, introduced in the same section.

This algorithm requires the state dynamics introduced in \eqref{eq::longitudinal_dyn_model}, \eqref{eq::battery_model} to be discretized and transformed to spatial domain. For each grid point $k$ along the route:
\begin{equation}
\label{eq::state_equations_DP}
\begin{aligned}
v_{k+1}^2 &= v_k^2 + 2 \Delta d_k \cdot \biggl(\frac{F_{trc,k} - F_{road,k}(v_k)}{M} \biggr) \\
\xi_{k+1}  &= \xi_k - \frac{\Delta d_k}{\bar{v}_{k}}\cdot \frac{\bar{I}_{batt,k}}{C_{nom}}, \quad \forall k = 1,\dots,N-1
\end{aligned}
\end{equation}
where $\Delta d_k$ is the distance over a stage (i.e. $\Delta d_k = d_{k+1}-d_k$, with $d_k$ as the distance traveled along the route at grid point $k$) and $\bar{v}_k \left(= \frac{v_k + v_{k+1}}{2} \right)$ is the average velocity over one stage. In this formulation, the signal phase information of each traffic light is deterministically incorporated as part of the initialization process before the trip begins. Varying timing information (i.e. time in each phase) however, cannot be utilized in the full-route optimization routine.

In this section, the DP-ECMS algorithm is introduced for the global optimization comprising of $N$ discrete grid points. In this methodology, a constant equivalence factor is used over the entire driving mission. The constraints of the $N$-step optimization are:
\begin{equation}
\label{eq::prb_constr_opt}
\begin{aligned}
v_{k} &\in \left[v_{k}^{min}, v_{k}^{max} \right], \quad \forall k = 2, \dots, N \\
\xi_k &\in \left[\xi^{min}, \xi^{max} \right], \quad \forall k = 2, \dots, N \\
v_1 &= v_1^{min}, \quad \xi_1 \in \left[\xi^{min}, \xi^{max} \right] \\
a_k &\in \left[a^{min}, a^{max} \right], \quad \forall k = 1, \dots, N \\
T_{eng,k} &\in \left[T_{eng,k}^{min}\left(v_k \right), T_{eng,k}^{max}\left(v_k \right) \right], \forall k = 1, \dots, N-1 \\
T_{bsg,k} &\in \left[T_{bsg,k}^{min}\left(v_k \right), T_{bsg,k}^{max}\left(v_k \right) \right], \forall k = 1, \dots, N-1
\end{aligned}
\end{equation}
where, $v_{veh}^{min}$ and $v_{veh}^{max}$ are the minimum and maximum route speed limits respectively; $\xi^{min}$ and $\xi^{max}$ represent the static limits applied on battery SoC variation; $T_{eng}^{min}$ and $T_{eng}^{max}$ are the state-dependent minimum and maximum torque limits of the engine respectively, and $T_{bsg}^{min}$ and $T_{bsg}^{max}$ are the state-dependent minimum and maximum BSG torque limits respectively. To ensure SoC-neutrality over the global optimization, a terminal constraint is applied on the battery SoC: $\xi_1 = \xi_N$. Further, dynamical constraints are imposed by the vehicle model dynamics, described in Equations \ref{eq::engine_model_T_eng_fuel} - \ref{eq::longitudinal_dyn_model}.

\begin{figure}[!htbp]
	\centering
	\includegraphics[width=\columnwidth]{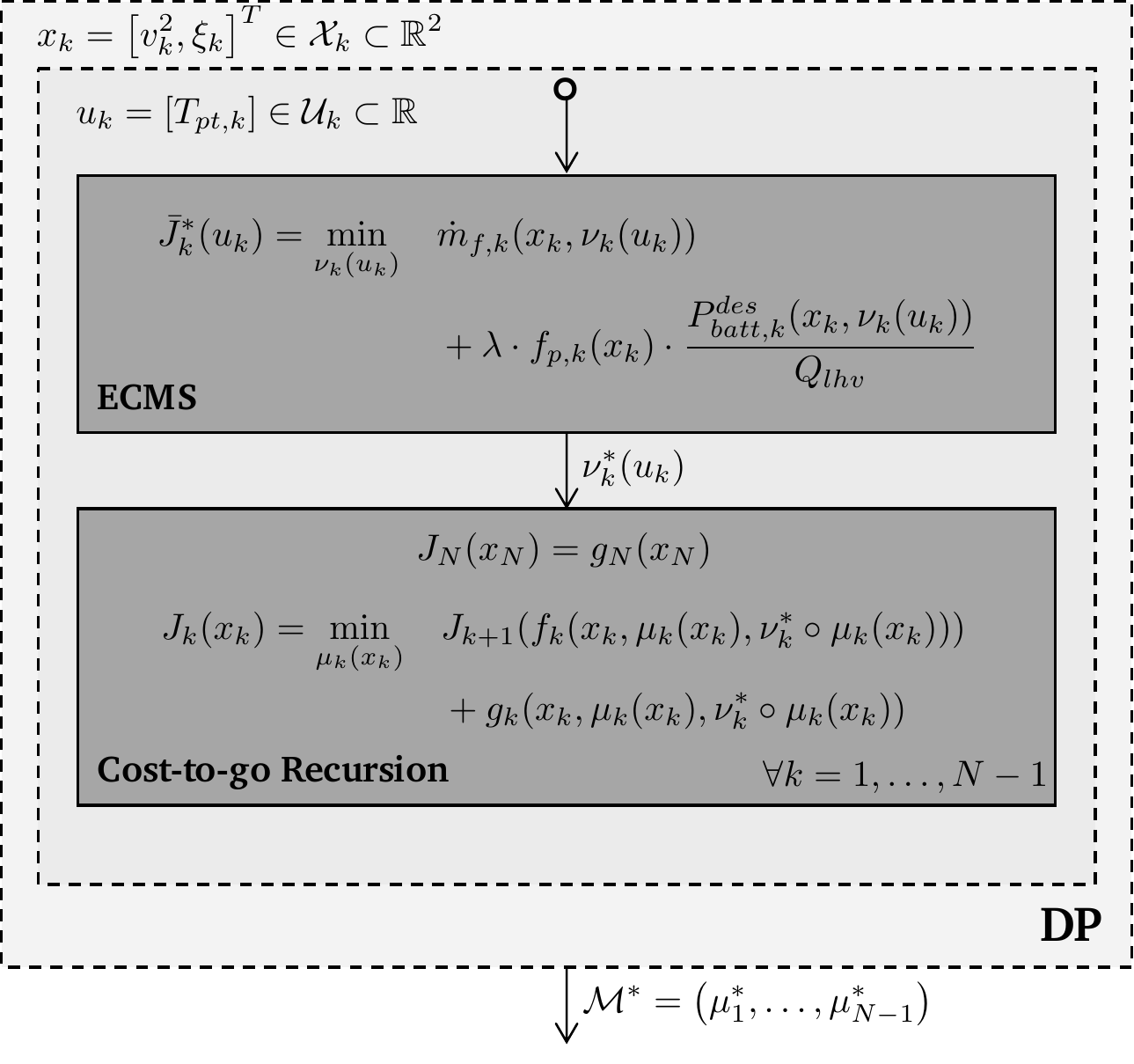}
	\caption{Schematic of the DP-ECMS algorithm for full-route optimization.}
	\label{fig::DPECMS_schematic}
\end{figure}

A key element in the DP-ECMS method is the mapping between the DP and the ECMS optimization problems. For a given powertrain torque (control input from DP), the ECMS yields the optimal torque split. Define at a grid point $k$, $\nu_k : u_k \to \bar{u}_k$ such that $\bar{u}_k^* = \nu_k^*(u_k)$ is the optimal BSG torque from the ECMS computed for each powertrain torque $u_k$. This map is used by the following DP problem, starting with terminal cost $J_N(x_N) = g_N(x_N)$:
\begin{equation}
\label{eq::DP_bellman_reformulated}
\begin{aligned}
J_k(x_k) = \min_{\mu_k(x_k)} \quad J_{k+1}(f_k(x_k,\mu_k(x_k), \nu_k^* \circ \mu_k(x_k))) &\\
+ g_k(x_k,\mu_k(x_k), \nu_k^* \circ \mu_k(x_k)), &\\
\forall k = 1,\dots,N-1 &
\end{aligned}
\end{equation}
where the $\circ$ operator denotes the composition of respective functions. In the model developed, the stage cost and cost-to-go functions are defined using the states and the torque split. As the control $u_k = \mu_k(x_k)$ contains only the powertrain torque, it is necessary to augment \eqref{eq::bellman} with $\nu_k^*(u_k)$ from the ECMS, that is obtained by solving the following instantaneous minimization problem at each stage $k$:
\begin{equation}
\label{eq::DPECMS_ECMS_eqn}
\begin{aligned}
\bar{J}_k^*(u_k) = \min_{\nu_k(u_k)} \quad &\dot{m}_{f,k}(x_k,\nu_k(u_k)) \\
&+ \lambda \cdot f_{pen,k}(x_k) \cdot \frac{P_{batt,k}^{des}(x_k,\nu_k(u_k))}{Q_{lhv}}
\end{aligned}
\end{equation}
Equations \ref{eq::DP_bellman_reformulated} and \ref{eq::DPECMS_ECMS_eqn} provide the optimal solution to the problem formulated in Section \ref{sec::problem_reformulation}. The DP-ECMS algorithm is compactly represented using the flowchart in Fig. \ref{fig::DPECMS_schematic}. At each grid point $k$ along the route, the DP explores all feasible combinations of the discretized state $x_k$ and control $u_k$ grids. At each discretized powertrain torque, the embedded ECMS algorithm determines the optimal torque split that is used in the DP cost-to-go recursion. Starting from the terminal stage (destination), the backward recursion yields the closed-loop optimal control policy as well as the corresponding optimal state trajectories over the full route.

As evident from the formulation, this approach utilizes a constant $\lambda$ over the entire driving mission. Additional calibration effort is necessary to appropriately select both $\lambda$ and the penalty function $f_{pen,k}$ that ensures boundedness of the SoC state:
\begin{equation}
\lambda \cdot f_{pen,k}(\xi_k) = \lambda_0 + \tan(-(\xi_k - \xi^{des})\cdot \lambda_1)
\end{equation}
where $\xi^{des}$ is the target value that is commonly chosen as the initial SoC $\xi_1$, and $(\lambda_0,\lambda_1)$ are constant calibration terms. Here, a $\tan(\cdot)$ function is chosen as its slope in the neighborhood of $\xi^{des}$ is nearly zero (can be tuned using $\lambda_1$). The optimal value of $\lambda_0$ that results in a charge-sustaining strategy is dependent on the route characteristics. It is tuned using the shooting method, in which the full-route optimization (in which the ECMS routine is contained) is run with different $\lambda_0$ to iteratively compute its optimal value.

\section{Look-Ahead Control using DP-ECMS}
\label{sec::DPECMS_grd_lambda_RH}

\begin{figure*}[!t]
	\centering
	\includegraphics[width=1.9\columnwidth]{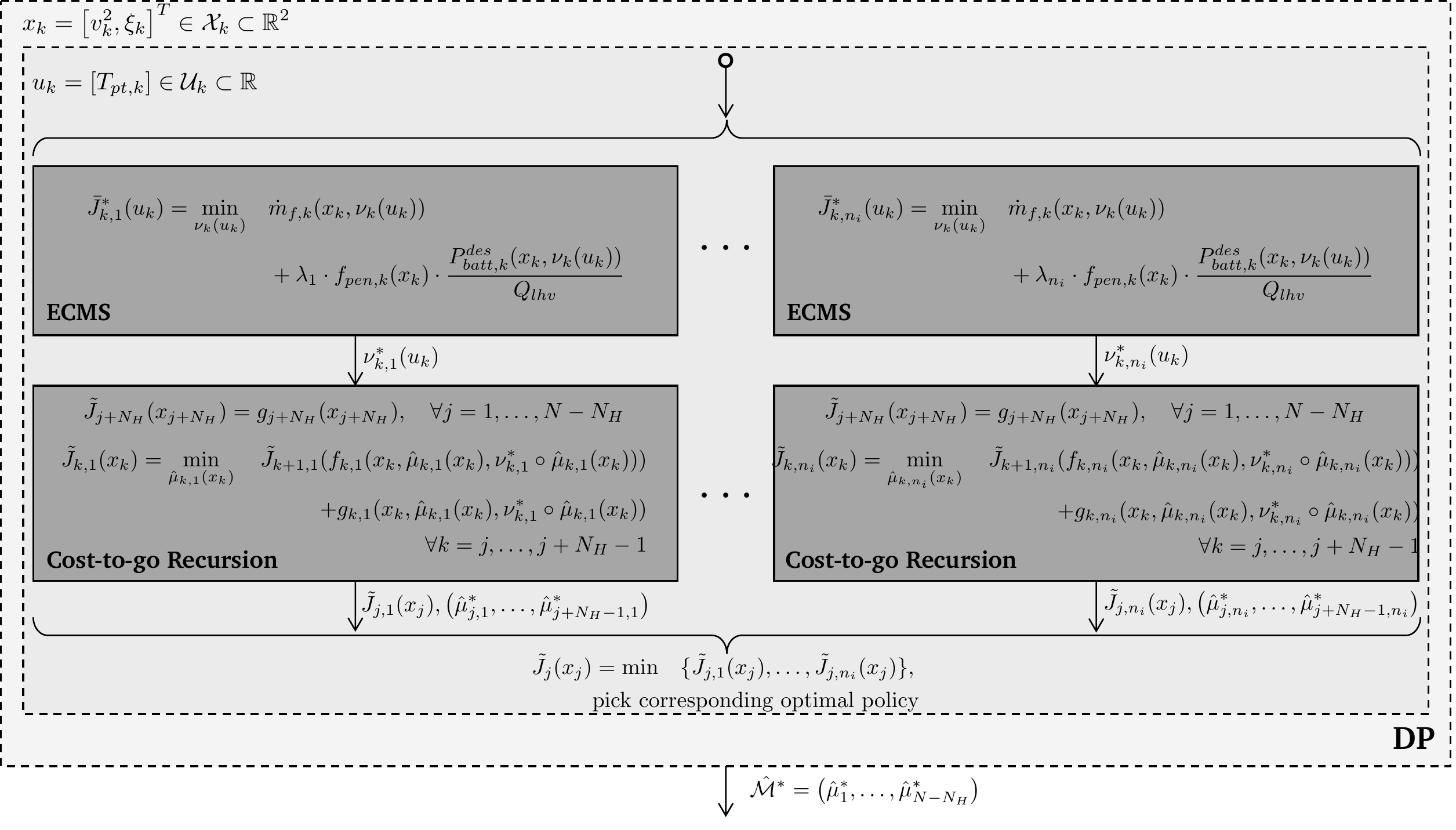}
	\caption{Schematic of the DP-ECMS algorithm for look-ahead control.}
	\label{fig::DPECMS_RH_schematic}
\end{figure*}

To accommodate variability in route conditions and/or uncertainty in route information in a computationally tractable manner, a look-ahead control scheme using DP-ECMS is proposed. Here, the full-route horizon of $N$-steps is truncated to $N_H \ll N$ steps. The goal of the receding horizon optimization in this paper is to add the capability of adapting to short-term changes in route conditions while yielding results comparable to the full-route optimization in the absence of such uncertainties or variabilities. To achieve this within the DP-ECMS framework, two key challenges need to be addressed: choice of terminal cost and constraints, and determination of an appropriate $\lambda$ for ECMS particularly for short horizons.

The terminal cost of the receding horizon optimization problem is constructed using principles in Approximate DP, specifically techniques based on approximation in the value space. For some valid map $\nu_k$, consider the following $N_H$-horizon one-step look-ahead control problem:
\begin{equation*}
\tilde{J}_{j+N_H}(x_{j+N_H}) = g_{j+N_H}(x_{j+N_H}), \quad \forall j = 1,\dots,N-N_H
\end{equation*}
\begin{equation}
\label{eq::DP_bellman_RH_approx}
\begin{aligned}
\tilde{J}_{k}(x_k) = \min_{\hat{\mu}_{k}(x_k)} \quad \tilde{J}_{k+1}(f_{k}(x_k,\hat{\mu}_{k}(x_k), \nu_{k}^* \circ \hat{\mu}_{k}(x_k))) &\\
+ g_{k}(x_k,\hat{\mu}_{k}(x_k), \nu_{k}^* \circ \hat{\mu}_{k}(x_k)), &\\
\quad \forall k = j,\dots,j+N_H-1 &
\end{aligned}
\end{equation}
where, $\tilde{J}_{j+N_H}$ (and as a result $\tilde{J}_{k+1}$) is the cost-to-go of a known suboptimal policy - a base policy, and $\hat{\mathcal{M}}^* := \left(\hat{\mu}_j^*, \dots, \hat{\mu}_{j+N_H-1}^* \right)$ is the rollout policy. This type of formulation, which falls under the class of rollout algorithms, offers the desirable property of cost improvement - as long as the base policy produces a feasible solution, the rollout algorithm also produces a feasible solution, whose cost is no worse than the cost corresponding to the base policy (proof in \cite{bertsekas1995dynamic}).

In this work, the base policy is assigned as the optimal cost-to-go function from the full-route DP optimization. Approximating the terminal cost in \eqref{eq::DP_bellman_RH_approx} as the optimal cost-to-go function from the full-route optimization results in \eqref{eq::DP_bellman_reformulated}, a derivative of the Bellman equation. In this way, while providing policy improvement and adaptation to short-term variabilities, this approach also provides a closed-loop optimal policy that matches the full-route optimization if minimal or no variabilities are experienced en route.

The methodology prescribed in Section \ref{sec::DPECMS_const_lambda_FR} for the full-route (or a long horizon) optimization involves defining a terminal constraint on the battery SoC and tuning the $\lambda$ for charge-sustaining behavior. Applying the same methodology to shorter horizons results in a highly conservative torque split strategy. One of the inherent benefits from adopting the aforementioned rollout algorithm approach is that it eliminates the need to define terminal constraints on the battery SoC, while ensuring approximately charge-sustaining behavior over the entire trip.

The DP-ECMS algorithm for look-ahead control is compactly represented using the flowchart in Fig. \ref{fig::DPECMS_RH_schematic}. In this approach, the optimal $\lambda$ for the ECMS algorithm is selected dynamically from a uniform $\lambda$ grid. For each value $\lambda_i$ in the grid, the following minimization problem is solved $\forall k = j,\dots,j+N_H-1, \quad \forall j = 1,\dots,N-N_H$:
\begin{equation}
\label{eq::DPECMS_grid}
\begin{aligned}
\bar{J}_{k,i}^*(u_k) = \min_{\nu_{k}(u_k)}& \quad \dot{m}_{f,k}(x_k,\nu_{k}(u_k)) \\
&+ \lambda_i \cdot f_{pen,k}(x_k) \cdot \frac{P_{batt,k}^{des}(x_k,\nu_{k}(u_k))}{Q_{lhv}} \\
& \qquad \qquad \qquad \qquad \qquad \forall i = 1,\dots,n_i 
\end{aligned}
\end{equation}
where $n_i$ is the number of equally spaced $\lambda$ candidates considered. To further reduce the computational cost of this instantaneous minimization, it is to be noted that terms $\dot{m}_{f,k}(x_k,\nu_{k}(u_k))$ and $\frac{P_{batt,k}^{des}(x_k,\nu_{k}(u_k))}{Q_{lhv}}$ remain the same $\forall i$. These terms are computed only once, and the cost $\bar{J}_{k,i}^*(u_k)$ is then obtained as a linear combination of these terms with only $\lambda_i$ changing for each $i$. The map $\nu_{k,i}^*(u_k)$ that defines the optimal torque split for each $u_k$ and $\lambda_i$ is used by the rollout algorithm below:
\begin{equation*}
\tilde{J}_{j+N_H}(x_{j+N_H}) = g_{j+N_H}(x_{j+N_H}), \quad \forall j = 1,\dots,N-N_H
\end{equation*}
\begin{equation}
\label{eq::DP_bellman_RH_DPECMS}
\begin{aligned}
\hat{J}_{k,i}(x_k) = \min_{\hat{\mu}_{k,i}(x_k)} \quad \hat{J}_{k+1,i}(f_{k,i}(x_k,\hat{\mu}_{k,i}(x_k), \nu_{k,i}^* \circ \hat{\mu}_{k,i}(x_k))) &\\
+ g_{k,i}(x_k,\hat{\mu}_{k,i}(x_k), \nu_{k,i}^* \circ \hat{\mu}_{k,i}(x_k)) &\\
\forall i = 1,\dots,n_i, \quad \forall k = j,\dots,j+N_H-1 &
\end{aligned}
\end{equation}
where $\left(\hat{\mu}_{j,i}^*, \dots, \hat{\mu}_{j+N_H-1,i}^* \right)$ is the rollout policy for each $\lambda_i$. This is subject to the same constraints defined in \eqref{eq::prb_constr_opt}, except that they are suitably truncated over the $N_H$ horizon. Finally, the $\lambda_i$ that results in the minimum cost-to-go approximate $\tilde{J}_j(x_j) $ (defined below) is selected as the optimal equivalence factor for that receding horizon.
\begin{equation*}
\tilde{J}_j(x_j) = \min \quad \{\tilde{J}_{j,1}(x_j),\dots,\tilde{J}_{j,n_i}(x_j) \}
\end{equation*}
The corresponding optimal control policy is applied as the torque split strategy. The highlights of this approach are that this structure still conforms to the assumptions made in the classical ECMS approach, in which the optimal $\lambda$ results in charge-sustaining behavior over the entire trip, and that the $\lambda$ exploration can be easily parallelized.

\section{Evaluation}
\label{sec::evaluation}

In this section, the different DP-ECMS approaches discussed in Sections \ref{sec::DPECMS_const_lambda_FR} and \ref{sec::DPECMS_grd_lambda_RH} are evaluated. Before the simulation results are presented, the selected test route and the benchmark solution developed are introduced.

\subsection{Test Route}

\begin{figure}[!htbp]
	\centering
	\includegraphics[width=0.9\columnwidth]{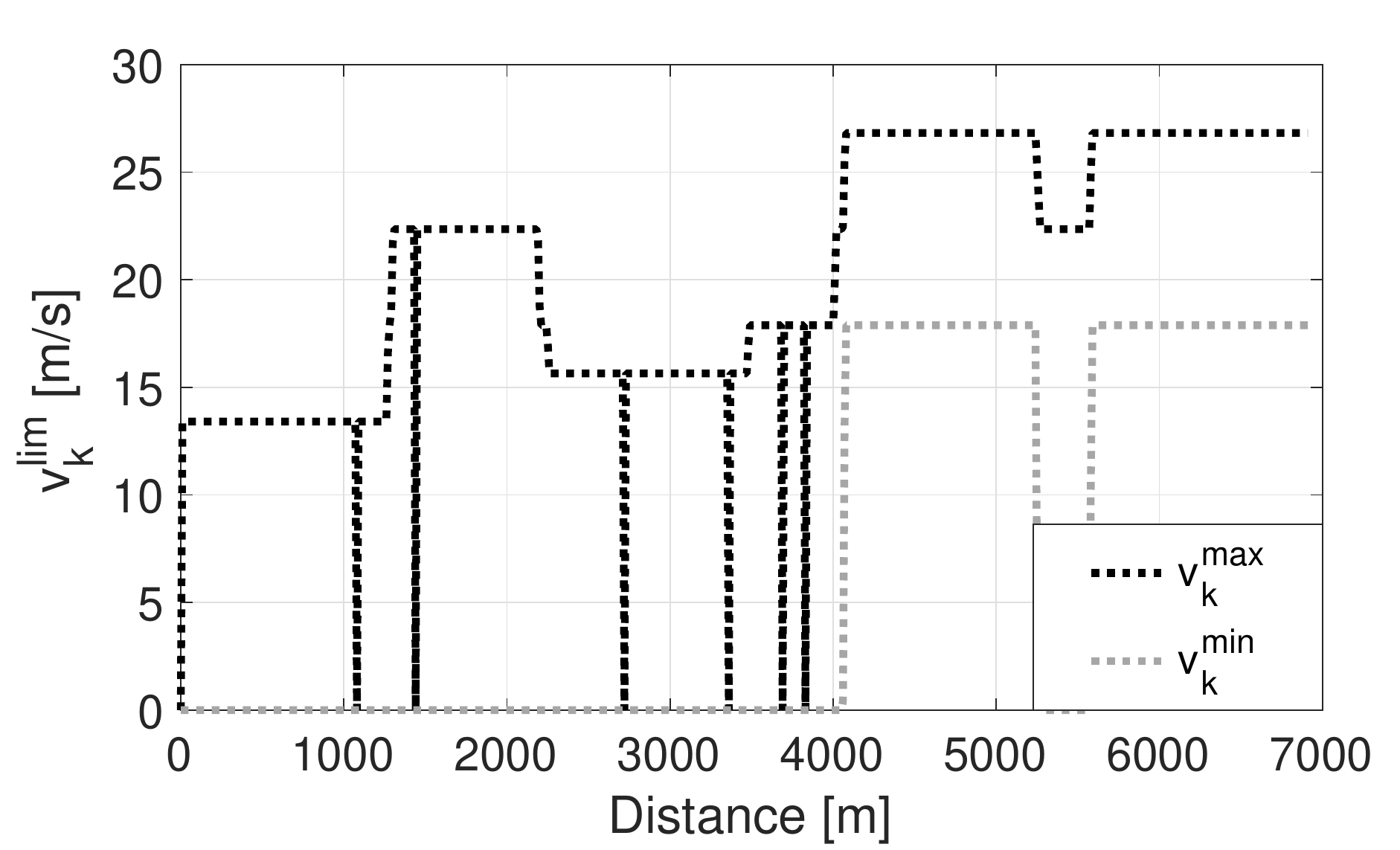}
	\caption{Speed limits and locations of stop signs along the test route.}
	\label{fig::R19_features}
\end{figure}

The test route over which the simulation study is performed is based on a mixed (urban/highway) route in the Columbus, OH region. This route is $\SI{7}{km}$ in length and comprises of 6 stop signs, as shown in Fig. \ref{fig::R19_features}.

\subsection{Benchmark Case}

The benchmark case developed considers the same problem formulated in Section \ref{sec::objective_problem}. A 2-state, 2-input DP is used to solve the full route optimization comprising of $N$ steps. Specifically, the state variables chosen are the vehicle velocity and the battery SoC: $x_{b,k} = \left[v_k, \xi_k \right]^\mathsf{T}, \quad \forall k = 1,\dots,N-1$. The engine torque and BSG torque are chosen as the control variables: $u_{b,k} = [T_{eng,k}, T_{bsg,k}]^\mathsf{T}$. For consistency, the vehicle and powertrain models used in this are identical to those used for developing the DP-ECMS method.

\subsection{Evaluation Metrics}

The optimization schemes developed are compared both qualitatively and quantitatively against the benchmark full-route DP optimization. For qualitative comparison, the resulting Pareto fronts from the multi-objective optimizations are compared, along with sample state trajectories. To quantitatively evaluate the performance of the DP-ECMS optimization routine, the cumulative cost that it incurs is compared with that incurred by the benchmark DP. The cumulative cost function used for comparison is shown below: 
\begin{equation*}
\begin{aligned}
J^*(x_1) &= \sum_{k = 1}^{N} \left(\gamma \cdot \frac{\dot{m}_{f,k}^*}{\dot{m}_f^{norm}} + (1-\gamma)\right) \cdot t_k^* \\
&= \gamma \cdot \frac{m_{f,N}^*}{\dot{m}_f^{norm}} + (1-\gamma) \cdot \tau_N^*
\end{aligned}
\end{equation*}
where $m_{f,N}^*,\tau_N^*$ are respectively the resulting cumulative fuel consumption and total travel time from applying the optimal control policy. For fair evaluation, the cumulative costs for both the DP-ECMS and benchmark DP algorithms are calculated while ensuring battery SoC neutrality over the entire driving mission.

\subsection{Evaluation of Full-Route DP-ECMS}

To recall, the full-route DP-ECMS refers to the approach in which the DP-ECMS algorithm is developed to solve the full-route optimization problem and the $\lambda$-tuning strategy comprises of computing a single $\lambda$ value over the entire trip using the shooting method. In this particular implementation of the full-route DP-ECMS, an unconstrained optimization routine is setup to tune the equivalence factor for achieving charge-sustaining behavior.

\begin{figure}[!htbp]
	\centering
	\includegraphics[width=0.9\columnwidth]{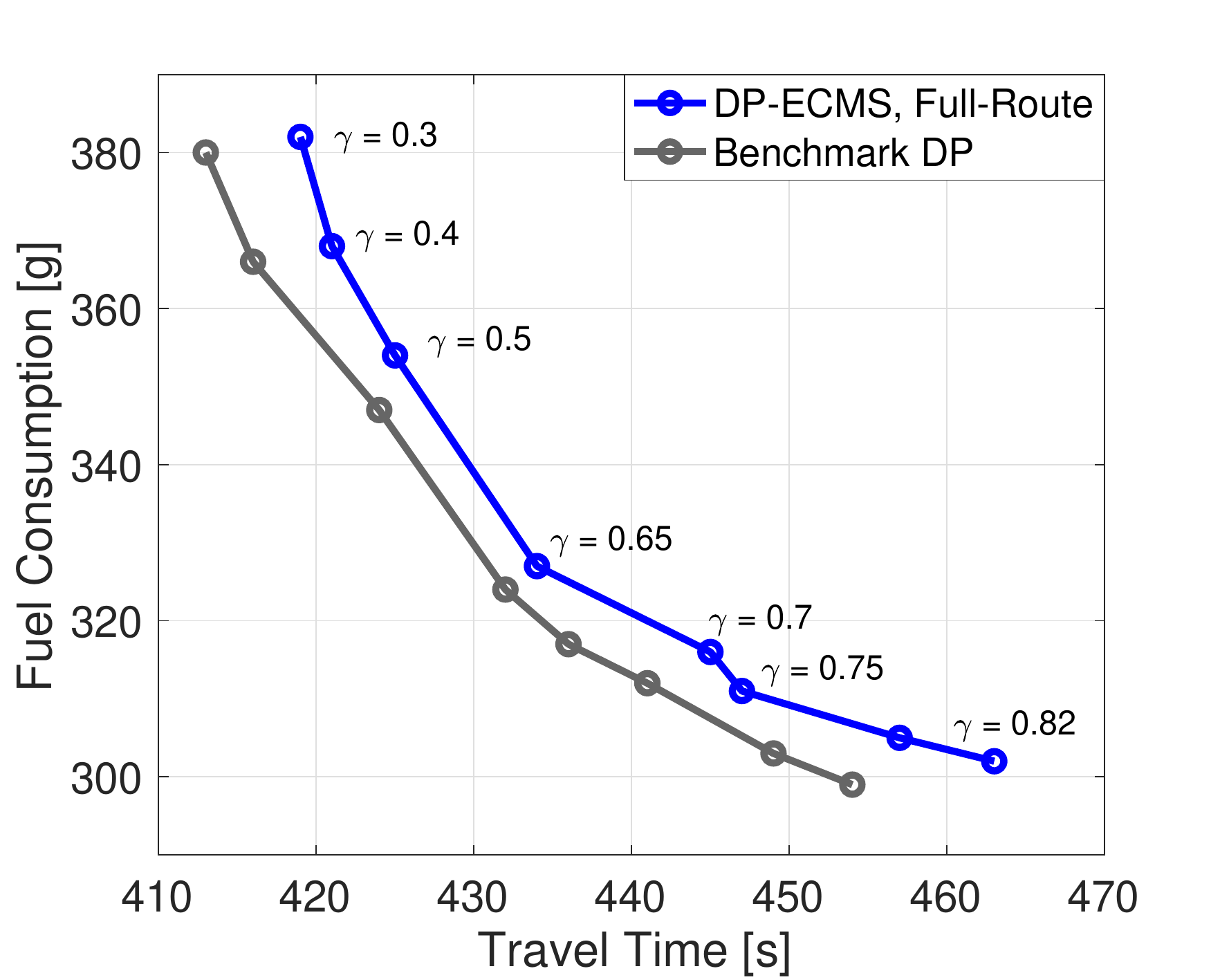}
	\caption{Pareto curve from DP-ECMS for full-route optimization and comparison with benchmark DP.}
	\label{fig::dp_ecms_full_route_pareto}
\end{figure}

Fig. \ref{fig::dp_ecms_full_route_pareto} shows sample results from the full-route DP-ECMS optimization over the mixed test route. A general note on the Pareto fronts is that lower values of $\gamma$ represent increasingly aggressive drivers, characterized by decreased travel time and increased fuel consumption. On comparison with the Pareto curve from the benchmark DP optimization, it is seen that the sub-optimal DP-ECMS optimization consumes only $\SIrange{0.5}{1}{\%}$ more fuel while taking $\SIrange{1.5}{2}{\%}$ longer to travel the same route.

\begin{figure}[!htbp]
	\centering
	\vspace{-3mm}
	\subfloat[Vehicle velocity and battery SoC profile, $\gamma = 0.65$]{\includegraphics[width=\columnwidth]{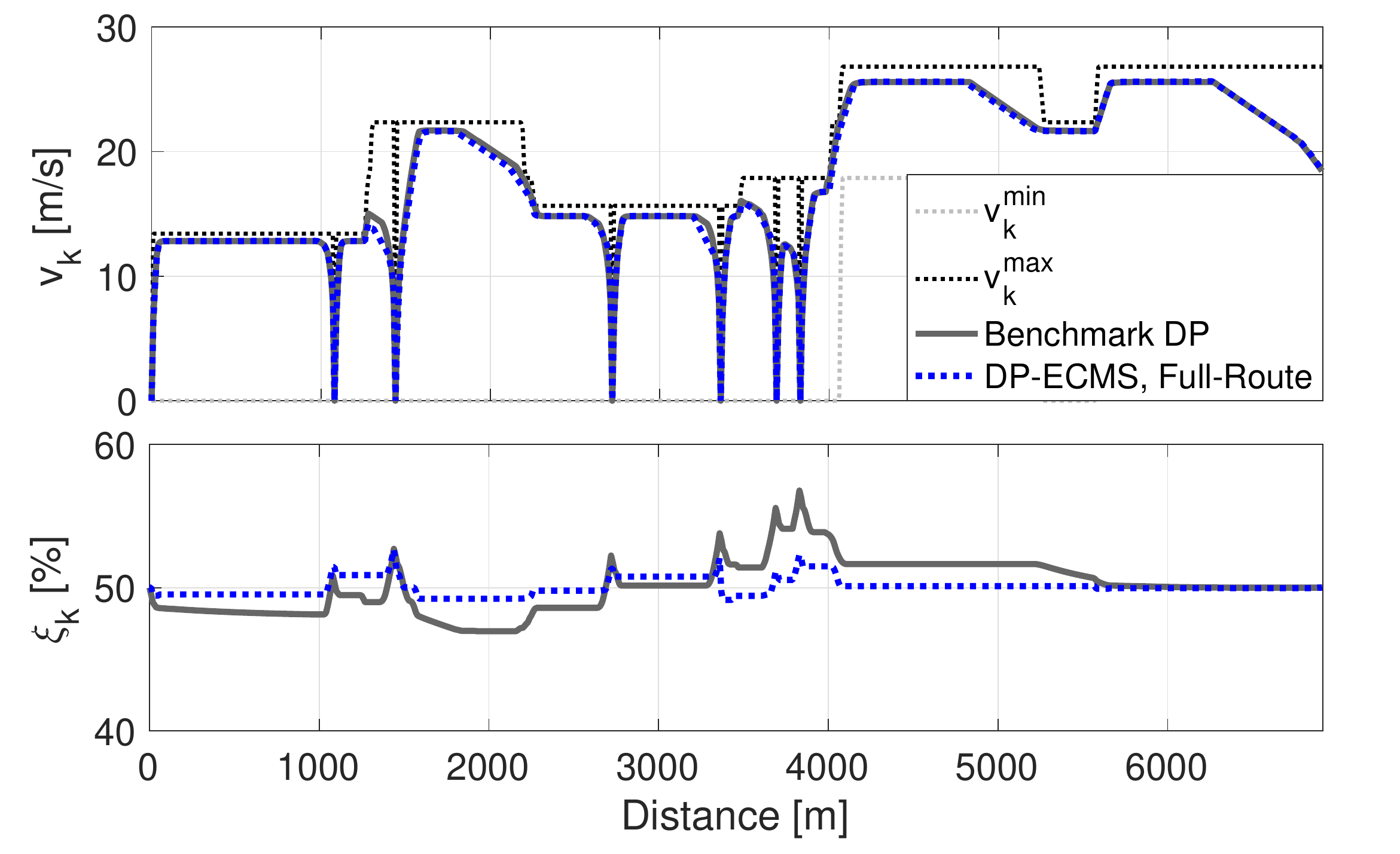}}
	\hfil
	\subfloat[Cumulative fuel consumption]{\includegraphics[width=0.7\columnwidth]{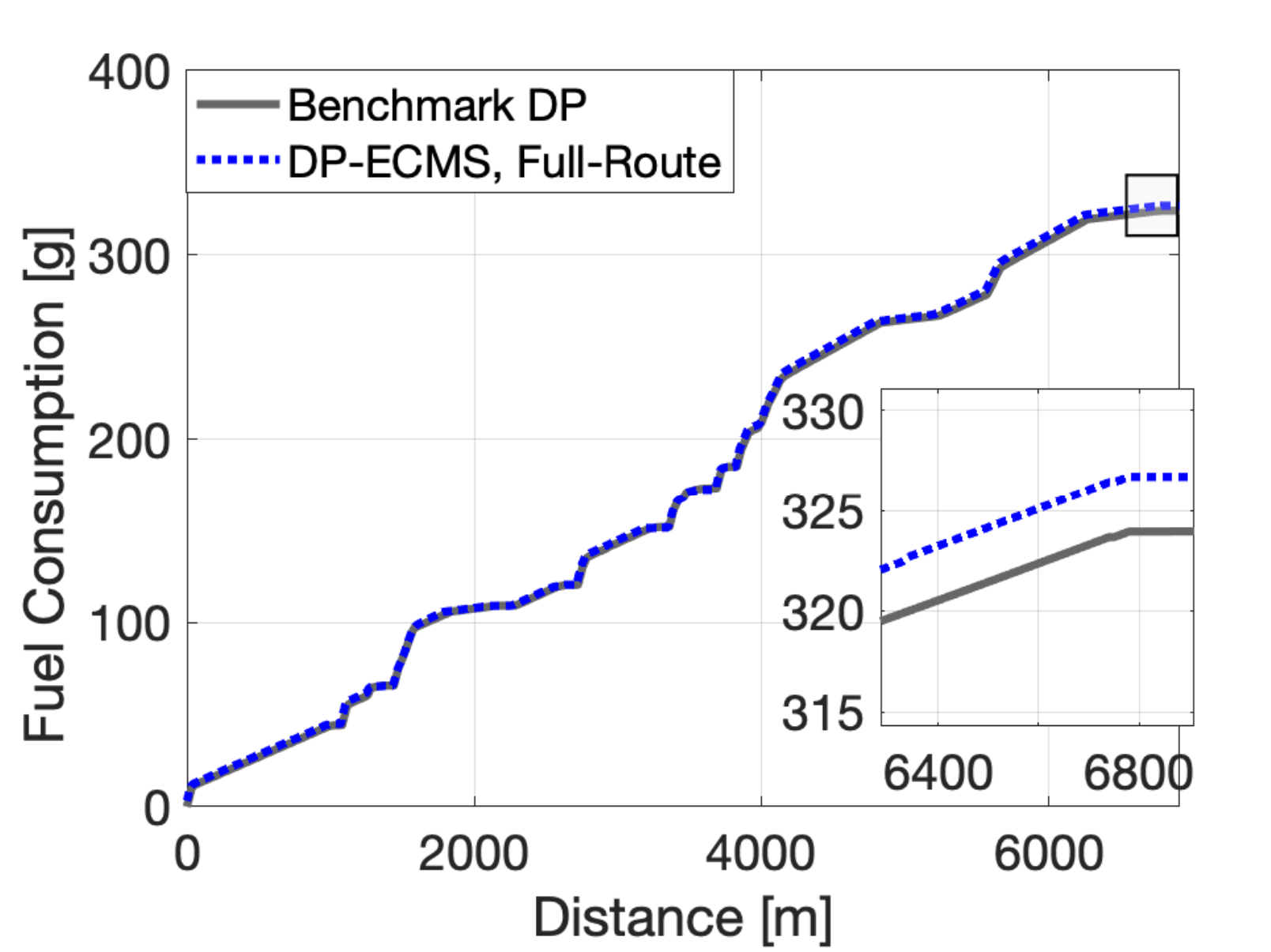}}
	\caption{Sample results from full-route optimization}
	\label{fig::dp_ecms_full_route_sample_results}
	\vspace{-1mm}
\end{figure}

In Fig. \ref{fig::dp_ecms_full_route_sample_results}(a), the resulting vehicle velocity and battery SoC profiles for ($\gamma = 0.65, \lambda_0 = 2.87$) are compared with the corresponding benchmark DP solution. While the vehicle velocity trajectory is nearly identical to that from the DP, the battery usage from the DP-ECMS control strategy is more conservative but charge sustaining nonetheless. In the selected case shown in Fig. \ref{fig::dp_ecms_full_route_sample_results}(b), the DP-ECMS routine consumes only $\SI{0.9}{\%}$ more fuel cumulatively than the benchmark over the entire driving mission.

\begin{table}[!htbp]
	\centering
	\begin{tabular}{cccc}
		\hline
		\multirow{2}{*}{$\gamma$} & Benchmark DP & DP-ECMS & \multirow{2}{*}{\begin{tabular}[c]{@{}c@{}}Cost Increment\\ $[\SI{}{\%}]$\end{tabular}} \\
		& $J^* [\SI{}{-}]$ & $J^* [\SI{}{-}]$ &  \\ \hline
		0.3 & 318  & 322 & +1.4 \\
		0.4 & 286 & 289 & +1.1 \\
		0.5 & 255 & 257 &  +0.5 \\
		0.65 & 204 & 205 & +0.6 \\
		0.7 & 186 & 189 & +1.4 \\
		0.75 & 168 & 170 & +0.8 \\
		0.8 & 150 & 152 & +1.3 \\
		0.82 & 143 & 145 & +1.6 
	\end{tabular}
	\caption{Quantitative evaluation of full-route optimization using DP-ECMS controller}
	\label{tab::DPECMS_FR_eval}
\end{table}

Table \ref{tab::DPECMS_FR_eval} summarizes the quantitative evaluation of the full-route DP-ECMS controller. The resulting cumulative cost is compared against that obtained from the benchmark DP optimization. For brevity, the argument of the cumulative cost is suppressed. The results from the proposed DP-ECMS controls are promising, as cost increments relative to the benchmark DP are always less than $\SI{2}{\%}$ over the entire mixed test route.

\subsection{Evaluation of Look-Ahead DP-ECMS}

To recall, the look-ahead DP-ECMS refers to the approach in which a look-ahead control problem is formulated to adapt to en route variabilities, and solved using the proposed DP-ECMS algorithm. Further, the dynamic $\lambda$-tuning strategy comprises of exploring a uniform $\lambda$ grid such that an optimal equivalence factor is selected for each receding horizon. In the look-ahead DP-ECMS implemented, to achieve faster computation times, the DP routines corresponding to each $\lambda_i$ in the grid are run in parallel.

\begin{figure}[!htbp]
	\centering
	\includegraphics[width=0.9\columnwidth]{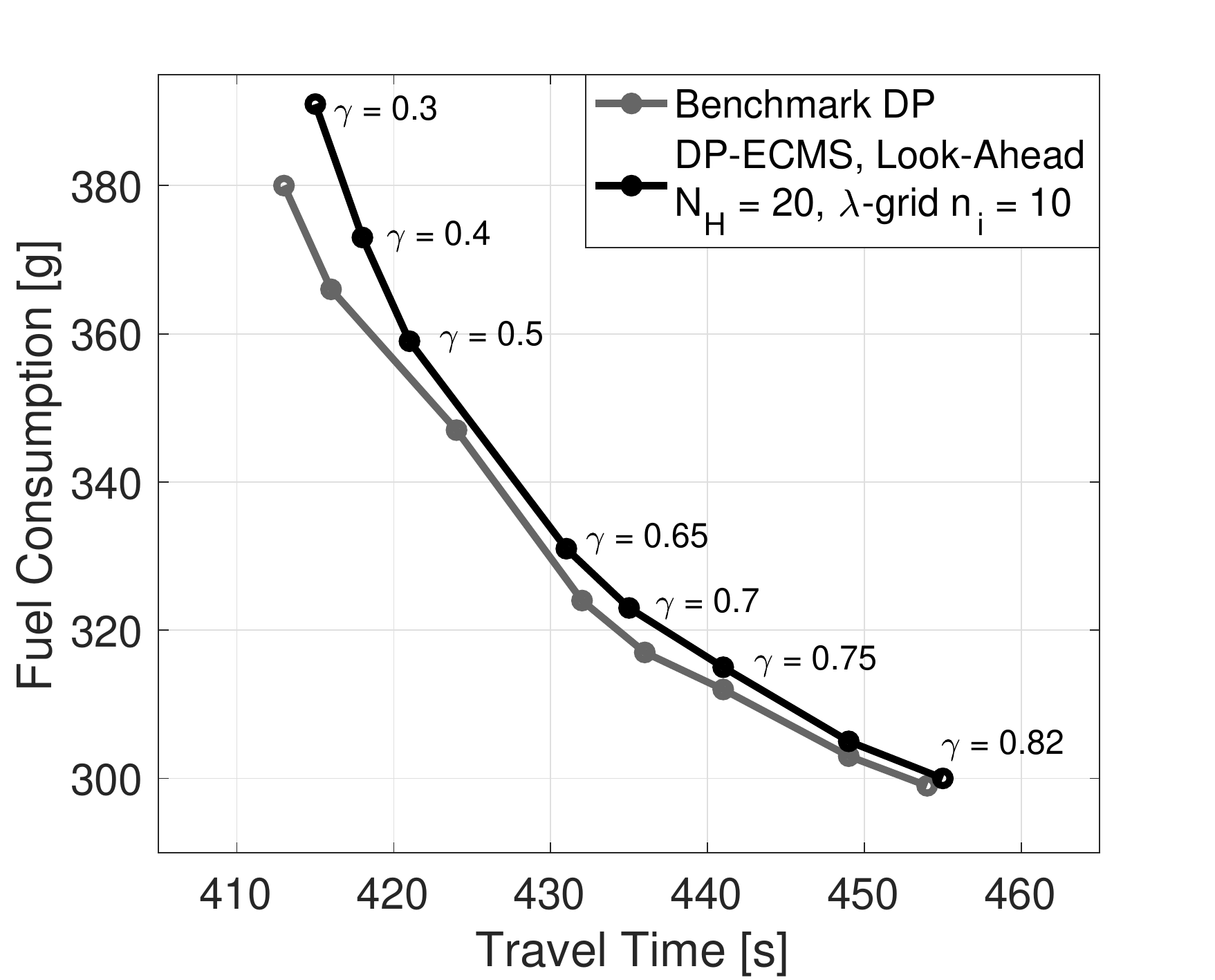}
	\caption{Pareto curve from DP-ECMS for look-ahead control and comparison with benchmark DP.}
	\label{fig::dp_ecms_look_ahead_pareto}
\end{figure}

Fig. \ref{fig::dp_ecms_look_ahead_pareto} shows the resulting Pareto front from the look-ahead DP-ECMS optimization over the mixed test route. Here, the equivalence factor is tuned dynamically from a $\lambda$-grid containing 10 equally spaced points. With a discretization step size of $\SI{10}{m}$ used in the DP-ECMS, a realistic choice is made for the horizon length, $N_H = 20$ (or $\SI{200}{m}$). On comparison with the Pareto curve from the benchmark DP, it is seen that the look-ahead DP-ECMS optimization consumes at most $\SI{2.5}{\%}$ more fuel while taking no more than $\SI{0.5}{\%}$ longer to travel the same route.

\begin{figure}[!htbp]
	\centering
	\vspace{-3mm}
	\subfloat[Vehicle velocity, battery SoC profile and dynamic equivalence factor, $\gamma = 0.65$]{\includegraphics[width=\columnwidth]{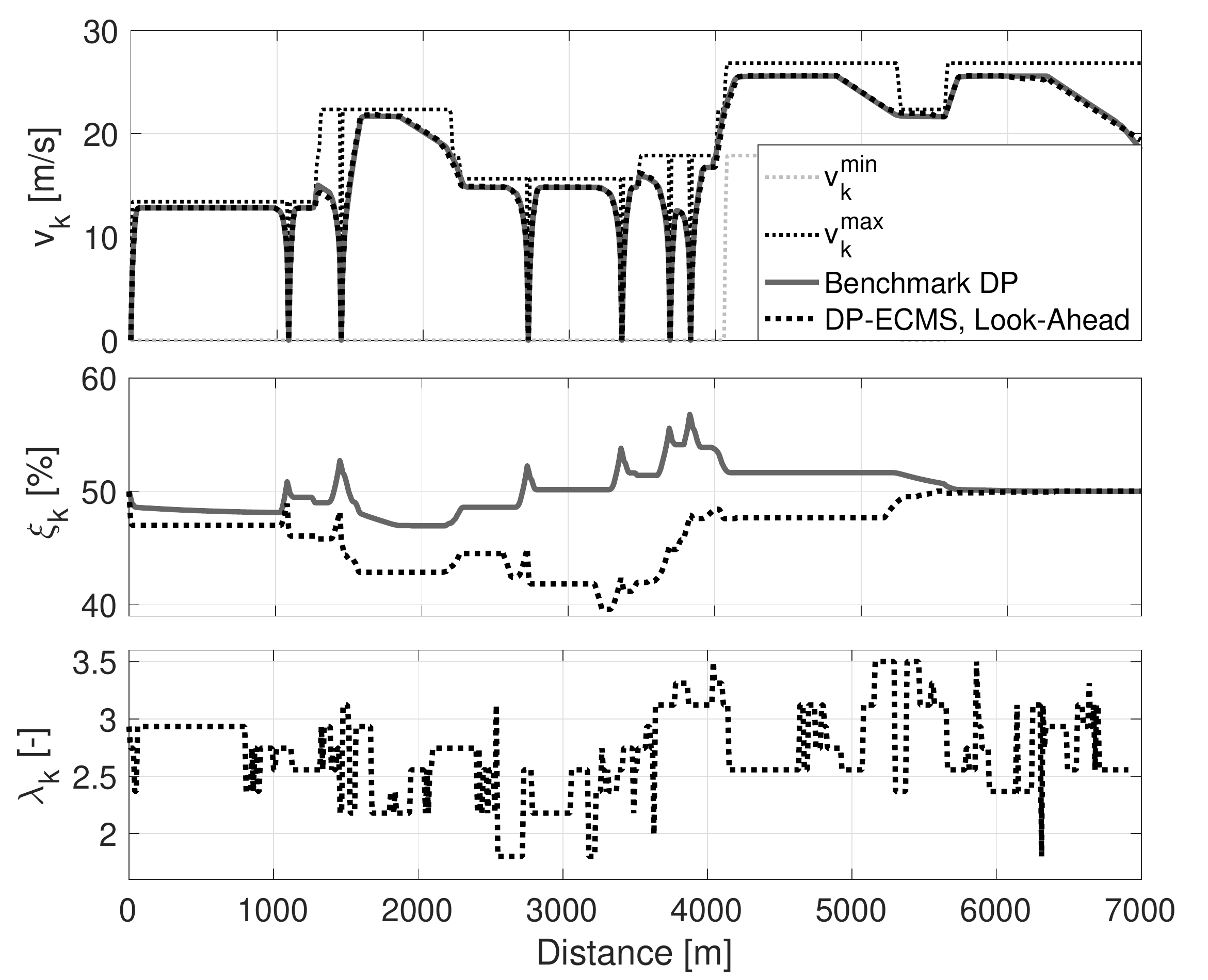}}
	\hfil
	\subfloat[Cumulative fuel consumption]{\includegraphics[width=0.7\columnwidth]{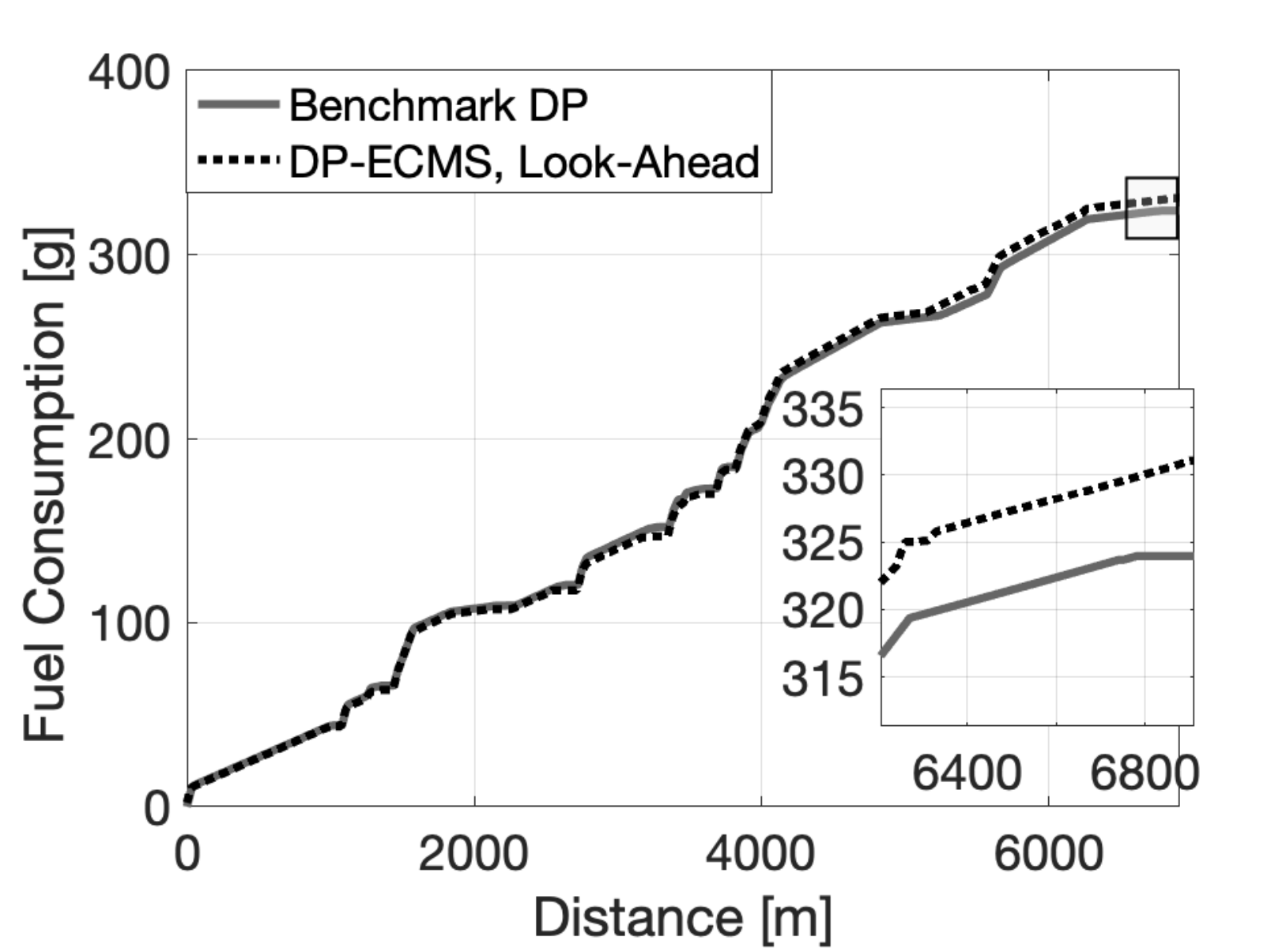}}
	\caption{Sample results from look-ahead control}
	\label{fig::dp_ecms_look_ahead_sample_results}
	\vspace{-1mm}
\end{figure}

Fig. \ref{fig::dp_ecms_look_ahead_sample_results}(a) shows the vehicle velocity, battery SoC, and dynamic equivalence factor profiles for ($\gamma = 0.65$). The vehicle velocity trajectory is nearly identical to that from the benchmark DP. In contrast to Fig. \ref{fig::dp_ecms_full_route_sample_results}(b), it is seen that the look-ahead DP-ECMS controller utilizes the battery to a greater extent, resulting in larger swings of the SoC. As desired, the torque split strategy is nominally charge-sustaining over the entire trip. Further, the behavior of the dynamic equivalence factor is as expected, where larger $\lambda$ values make battery use more expensive. In the selected case shown in Fig. \ref{fig::dp_ecms_look_ahead_sample_results}(b), the DP-ECMS routine consumes around $\SI{2}{\%}$ more fuel cumulatively than the benchmark over the entire driving mission.

\begin{table}[!htbp]
	\centering
	\begin{tabular}{ccccc}
		\hline
		\multirow{2}{*}{\begin{tabular}[c]{@{}c@{}}$\lambda$-Grid\\ $n_i$\end{tabular}} & \multirow{2}{*}{$\gamma$} & Benchmark DP & DP-ECMS & \multirow{2}{*}{\begin{tabular}[c]{@{}c@{}}Cost Increment\\ $[\SI{}{\%}]$\end{tabular}} \\
		&  & $J^* [\SI{}{-}]$ & $J^* [\SI{}{-}]$ &  \\ \hline
		\multirow{8}{*}{4} & 0.3 & 318 & 320 & +0.9 \\
		& 0.4 & 286 & 289 & +0.8 \\
		& 0.5 & 255 & 256 & +0.2 \\
		& 0.65 & 204 & 205 & +0.6 \\
		& 0.7 & 186 & 188 & +0.6 \\
		& 0.75 & 168 & 170 & +0.4 \\
		& 0.8 & 150 & 152 & +0.8 \\
		& 0.82 & 143 & 144 & +0.6 \\ \hline
		\multirow{8}{*}{10} & 0.3 & 318 & 320 & +0.7 \\
		& 0.4 & 286 & 288 & +0.7 \\
		& 0.5 & 255 & 255 & +0 \\
		& 0.65 & 204 & 205 & +0.4 \\
		& 0.7 & 186 & 187 & +0.4 \\
		& 0.75 & 168 & 169 & +0.3 \\
		& 0.8 & 150 & 151 & +0.3 \\
		& 0.82 & 143 & 143 & +0.3 \\ \hline
		\multirow{8}{*}{40} & 0.3 & 318 & 319 & +0.6 \\
		& 0.4 & 286 & 287 & +0.5 \\
		& 0.5 & 255 & 255 & +0 \\
		& 0.65 & 204 & 205 & +0.3 \\
		& 0.7 & 186 & 187 & +0.4 \\
		& 0.75 & 168 & 169 & +0.3  \\
		& 0.8 & 150 & 150 & +0.1 \\
		& 0.82 & 143 & 143 & +0
	\end{tabular}
\caption{Quantitative evaluation of look-ahead control using DP-ECMS}
\label{tab::DPECMS_RH_eval}
\end{table}

Table \ref{tab::DPECMS_RH_eval} summarizes the quantitative evaluation of the look-ahead DP-ECMS controller. Further, the impact of $\lambda$ grid size (i.e. number of potential equivalence factor options) on the optimization results is evaluated by considering three different sizes $n_i = \{4,10,40\}$. The benchmark DP solution for each of these cases remains unchanged.

The cumulative cost from the look-ahead DP-ECMS is now compared against that obtained from the benchmark DP. It is seen that the cost increments relative to the benchmark DP optimization are further reduced compared to the constant $\lambda$ approach over the full route. Here, the look-ahead DP-ECMS scheme is always less than $\SI{1}{\%}$ sub-optimal over the entire mixed test route. These results suggest that being able to adapt the equivalence factor en route can bring noticeable improvement in both the battery usage as well as the ``proximity to optimality''.

\subsection{A Note on Computational Effort and Choice of Horizon Length}

The search spaces of the benchmark DP and the DP-ECMS algorithms can be used as intuitive measures of their respective time and space complexity.


The complexity of the DP-ECMS optimization method, with respect to the benchmark DP formulation with computational complexity \eqref{eq::computation_DP}, is given by:
\begin{equation}
\label{eq::computation_DP_ECMS}
\tilde{n}_{c} = \mathcal{O} \left( N \cdot \prod_{i = 1}^{n} \tilde{n}_{x,i}  \cdot \prod_{i = 1}^{m-1} \tilde{n}_{u,i} \cdot \tilde{n}_{\bar{u}}\right)
\end{equation}
where $N$ is the number of stages, $\tilde{n}_x$ and $\tilde{n}_u$ are the number of discretized points in the state and control grids respectively and $\tilde{n}_{\bar{u}}$ refers to the discretization of the ECMS control (effectively the candidates for determining the torque split). Note that the dimension of the input space is now reduced by one since the DP-ECMS only explores over different powertrain torque values, with the power split being optimized within the ECMS. Here, the number of computations is further reduced as a smaller portion of the model is evaluated in simulating the ECMS controller. Further, a significant part of the model evaluations can be reused when performing the $\lambda$ grid search over different equivalence factors in \eqref{eq::DPECMS_grid}, thereby not significantly increasing the computational complexity.

When a vanilla DP algorithm is used to determine the optimal torque split strategy, the discretization of the $\xi_k$ grid is quite dense to minimize interpolation errors and avoid infeasibilities. In the DP-ECMS, the sole purpose of retaining the SoC as a state is to ensure constraint satisfaction over the optimization horizon. Simulation studies have revealed that the granularity of the $\xi_k$ has minimal impact on the actual torque split optimization as this is handled by the embedded ECMS routine. Thus, the proposed structure provides the freedom to select a highly coarsened $\xi_k$ grid without sacrificing on optimality and constraint satisfaction.

Simulations have shown that the ECMS allows for having a coarser discretization $n_{\bar{u}}$ compared to $n_u$. Further, these computations are much simpler in nature (with very few, low-dimensional interpolations). As a result, depending on the system configuration and application, the DP-ECMS optimization routine can offer over $10\times$ reduction in the number of computations performed.

\begin{figure}[!htbp]
	\centering
	\includegraphics[width=0.9\columnwidth]{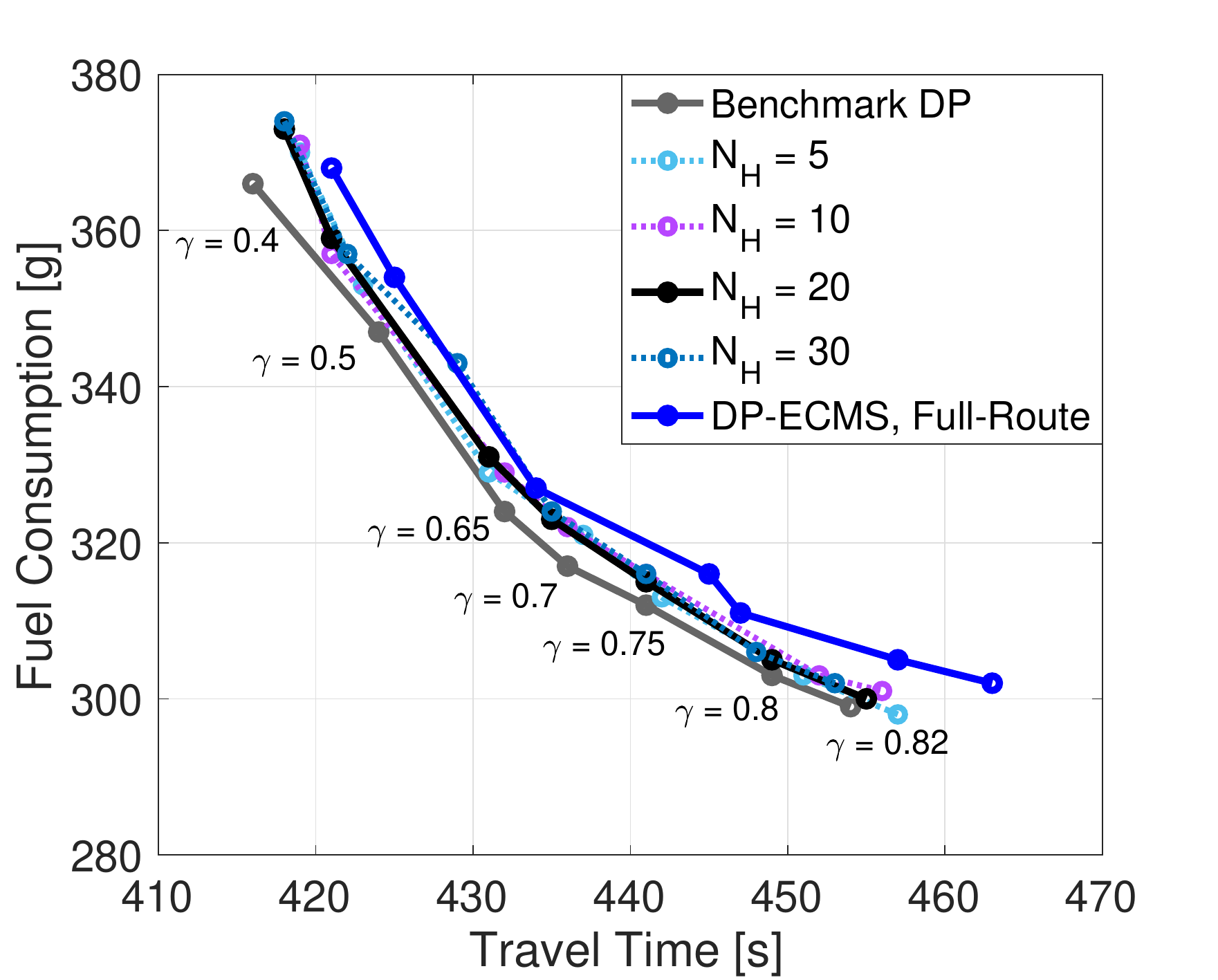}
	\caption{Pareto curve from DP-ECMS for look-ahead optimization and comparison with benchmark DP.}
	\label{fig::dp_ecms_look_ahead_pareto_N_H_study}
\end{figure}

Fig. \ref{fig::dp_ecms_look_ahead_pareto_N_H_study} shows the impact of horizon length on the optimality of the resulting solution. Here, the Pareto curves corresponding to $N_H = \{5,10,20,30 \}$ are all enclosed within the Pareto curves from the benchmark DP and full-route DP-ECMS. This observation becomes fairly obvious by recognizing that the performance of the benchmark DP cannot be mathematically improved, and by considering the upper limit of $N_H$, that eventually spans the entire (or remaining) route. As the value function of the full-route optimization is used as the terminal cost of the look-ahead control, it is expected that shorter $N_H$ will result in a Pareto front that more closely matches the benchmark DP. From these results, a choice of $N_H = 20$ (as made in Fig. \ref{fig::dp_ecms_look_ahead_pareto}) is justified as a reasonable compromise between adaptability to variabilities en route while preserving information from the original full-route DP and computational cost incurred. However, it is to be noted that this choice of horizon length cannot be generalized. Some of these observations are subject to change, depending on the powertrain configuration, and features of the test route i.e. speed limits, density of stop signs, elevation and so on.

\section{Conclusion}
\label{sec::conclusion}

This paper proposes a multi-layer hierarchical control framework, termed DP-ECMS, for HEV eco-driving applications. The proposed control architecture uses GPS and route information to jointly compute the optimal vehicle velocity and powertrain torque split trajectories with the aims to significantly improve the energy economy while providing a pathway to real-time implementation in a vehicle. Embedding the ECMS algorithm within the DP framework offers computational benefits without appreciable compromise in the performance of the optimization.

To accommodate en route variabilities and/or uncertainty in route information in a computationally tractable manner, a look-ahead optimization scheme using the DP-ECMS algorithm is developed. Here, the dynamic program is solved using the rollout algorithm, a technique based on approximation in the value space. The corresponding $\lambda$-tuning strategy evaluates multiple $\lambda$ values in parallel to pick the optimal candidate that results in the minimum cost-to-go approximate. Here, it is shown that the proposed technique results in nominally charge-sustaining behavior over the entire trip.

Relative to the benchmark DP optimization, it is seen that the cost increments from the full-route DP-ECMS consumes are always less than $\SI{2}{\%}$ over the evaluated mixed route. Further, the look- ahead DP-ECMS algorithm is consistently less than $\SI{1}{\%}$ sub-optimal over different $\lambda$ grid sizes. While the choice of the horizon length and route features are recognized to impact the results obtained, the proposed DP-ECMS method significantly reduces the computational costs incurred in the optimization, and is a promising solution for real-time HEV eco-driving applications.


%

\section*{Acknowledgment}

Our team gratefully acknowledges the support from the United States Department of Energy, Advanced Research Projects Agency – Energy (award number DE-AR0000794). The authors are grateful to Mr. Leo Bauer for the discussions and research efforts as a part of his Master's thesis \cite{bauer2018distance}, which eventually matured into this work.

\ifCLASSOPTIONcaptionsoff
  \newpage
\fi



%
\bibliographystyle{ieeetr}
\bibliography{references} 

\begin{thebibliography}{10}

\bibitem{ozatay2014cloud}
E.~Ozatay, S.~Onori, J.~Wollaeger, U.~Ozguner, G.~Rizzoni, D.~Filev,
  J.~Michelini, and S.~Di~Cairano, ``Cloud-based velocity profile optimization
  for everyday driving: A dynamic-programming-based solution,'' {\em IEEE
  Transactions on Intelligent Transportation Systems}, vol.~15, no.~6,
  pp.~2491--2505, 2014.

\bibitem{grumert2015analysis}
E.~Grumert, X.~Ma, and A.~Tapani, ``Analysis of a cooperative variable speed
  limit system using microscopic traffic simulation,'' {\em Transportation
  research part C: emerging technologies}, vol.~52, pp.~173--186, 2015.

\bibitem{jin2016power}
Q.~Jin, G.~Wu, K.~Boriboonsomsin, and M.~J. Barth, ``Power-based optimal
  longitudinal control for a connected eco-driving system,'' {\em IEEE
  Transactions on Intelligent Transportation Systems}, vol.~17, no.~10,
  pp.~2900--2910, 2016.

\bibitem{sciarretta2015optimal}
A.~Sciarretta, G.~De~Nunzio, and L.~Ojeda, ``Optimal ecodriving control:
  Energy-efficient driving of road vehicles as an optimal control problem,''
  {\em IEEE Control Systems Magazine}, vol.~35, no.~5, pp.~71--90, 2015.

\bibitem{wan2016optimal}
N.~Wan, A.~Vahidi, and A.~Luckow, ``Optimal speed advisory for connected
  vehicles in arterial roads and the impact on mixed traffic,'' {\em
  Transportation Research Part C: Emerging Technologies}, vol.~69,
  pp.~548--563, 2016.

\bibitem{mensing2013trajectory}
F.~Mensing, E.~Bideaux, R.~Trigui, and H.~Tattegrain, ``Trajectory optimization
  for eco-driving taking into account traffic constraints,'' {\em
  Transportation Research Part D: Transport and Environment}, vol.~18,
  pp.~55--61, 2013.

\bibitem{dib2014optimal}
W.~Dib, A.~Chasse, P.~Moulin, A.~Sciarretta, and G.~Corde, ``Optimal energy
  management for an electric vehicle in eco-driving applications,'' {\em
  Control Engineering Practice}, vol.~29, pp.~299--307, 2014.

\bibitem{hellstrom2009look}
E.~Hellstr{\"o}m, M.~Ivarsson, J.~{\AA}slund, and L.~Nielsen, ``Look-ahead
  control for heavy trucks to minimize trip time and fuel consumption,'' {\em
  Control Engineering Practice}, vol.~17, no.~2, pp.~245--254, 2009.

\bibitem{hellstrom2010design}
E.~Hellstr{\"o}m, J.~{\AA}slund, and L.~Nielsen, ``Design of an efficient
  algorithm for fuel-optimal look-ahead control,'' {\em Control Engineering
  Practice}, vol.~18, no.~11, pp.~1318--1327, 2010.

\bibitem{ozatay2017velocity}
E.~Ozatay, U.~Ozguner, and D.~Filev, ``Velocity profile optimization of on road
  vehicles: Pontryagin's maximum principle based approach,'' {\em Control
  Engineering Practice}, vol.~61, pp.~244--254, 2017.

\bibitem{uebel2017optimal}
S.~Uebel, N.~Murgovski, C.~Tempelhahn, and B.~B{\"a}ker, ``Optimal energy
  management and velocity control of hybrid electric vehicles,'' {\em IEEE
  Transactions on Vehicular Technology}, vol.~67, no.~1, pp.~327--337, 2017.

\bibitem{olin2019reducing}
P.~Olin, K.~Aggoune, L.~Tang, K.~Confer, J.~Kirwan, S.~R. Deshpande, S.~Gupta,
  P.~Tulpule, M.~Canova, and G.~Rizzoni, ``Reducing fuel consumption by using
  information from connected and automated vehicle modules to optimize
  propulsion system control,'' tech. rep., SAE Technical Paper, 2019.

\bibitem{heppeler2014fuel}
G.~Heppeler, M.~Sonntag, and O.~Sawodny, ``Fuel efficiency analysis for
  simultaneous optimization of the velocity trajectory and the energy
  management in hybrid electric vehicles,'' {\em IFAC Proceedings Volumes},
  vol.~47, no.~3, pp.~6612--6617, 2014.

\bibitem{sun2014velocity}
C.~Sun, X.~Hu, S.~Moura, and F.~Sun, ``Velocity predictors for predictive
  energy management in hybrid electric vehicles,'' {\em IEEE Transactions on
  Control Systems Technology}, vol.~23, no.~3, pp.~1197--1204, 2014.

\bibitem{xiang2017energy}
C.~Xiang, F.~Ding, W.~Wang, and W.~He, ``Energy management of a dual-mode
  power-split hybrid electric vehicle based on velocity prediction and
  nonlinear model predictive control,'' {\em Applied energy}, vol.~189,
  pp.~640--653, 2017.

\bibitem{liu2017reinforcement}
T.~Liu, X.~Hu, S.~Li, and D.~Cao, ``Reinforcement learning optimized look-ahead
  energy management of a parallel hybrid electric vehicle,'' {\em IEEE/ASME
  Transactions on Mechatronics}, vol.~22, no.~4, pp.~1497--1507, 2017.

\bibitem{homchaudhuri2016fast}
B.~HomChaudhuri, A.~Vahidi, and P.~Pisu, ``Fast model predictive control-based
  fuel efficient control strategy for a group of connected vehicles in urban
  road conditions,'' {\em IEEE Transactions on Control Systems Technology},
  vol.~25, no.~2, pp.~760--767, 2016.

\bibitem{sun2018robust}
C.~Sun, J.~Guanetti, F.~Borrelli, and S.~Moura, ``Robust eco-driving control of
  autonomous vehicles connected to traffic lights,'' {\em arXiv preprint
  arXiv:1802.05815}, 2018.

\bibitem{lim2016distance}
H.~Lim, W.~Su, and C.~C. Mi, ``Distance-based ecological driving scheme using a
  two-stage hierarchy for long-term optimization and short-term adaptation,''
  {\em IEEE Transactions on Vehicular Technology}, vol.~66, no.~3,
  pp.~1940--1949, 2016.

\bibitem{heppeler2017predictive}
G.~Heppeler, M.~Sonntag, U.~Wohlhaupter, and O.~Sawodny, ``Predictive planning
  of optimal velocity and state of charge trajectories for hybrid electric
  vehicles,'' {\em Control Engineering Practice}, vol.~61, pp.~229--243, 2017.

\bibitem{paden2016survey}
B.~Paden, M.~{\v{C}}{\'a}p, S.~Yong, D.~Yershov, and E.~Frazzoli, ``A survey of
  motion planning and control techniques for self-driving urban vehicles,''
  {\em IEEE Transactions on intelligent vehicles}, vol.~1, no.~1, pp.~33--55,
  2016.

\bibitem{grumert2017using}
E.~Grumert and A.~Tapani, ``Using connected vehicles in a variable speed limit
  system,'' {\em Transportation Research Procedia}, vol.~27, pp.~85--92, 2017.

\bibitem{bellman1966dynamic}
R.~Bellman, ``Dynamic programming,'' {\em Science}, vol.~153, no.~3731,
  pp.~34--37, 1966.

\bibitem{guzzella2007vehicle}
L.~Guzzella, A.~Sciarretta, {\em et~al.}, {\em Vehicle propulsion systems},
  vol.~1.
\newblock Springer, 2007.

\bibitem{bellman1964dynamic}
R.~Bellman and R.~Karush, {\em Dynamic Programming: A Bibliography of Theory
  and Application}.
\newblock Memorandum (Rand Corporation), Rand Corporation, 1964.

\bibitem{Larson:1982:PDP:578147}
R.~E. Larson and J.~L. Casti, {\em Principles of Dynamic Programming: Advanced
  Theory and Applications}.
\newblock New York, NY, USA: Marcel Dekker, Inc., 1982.

\bibitem{paganelli2002equivalent}
G.~Paganelli, S.~Delprat, T.~Guerra, J.~Rimaux, and J.~Santin, ``Equivalent
  consumption minimization strategy for parallel hybrid powertrains,'' in {\em
  Vehicular Technology Conference. IEEE 55th Vehicular Technology Conference.
  VTC Spring 2002 (Cat. No. 02CH37367)}, vol.~4, pp.~2076--2081, IEEE, 2002.

\bibitem{musardo2005ecms}
C.~Musardo, G.~Rizzoni, Y.~Guezennec, and B.~Staccia, ``A-ecms: An adaptive
  algorithm for hybrid electric vehicle energy management,'' {\em European
  Journal of Control}, vol.~11, no.~4-5, pp.~509--524, 2005.

\bibitem{pisu2007comparative}
P.~Pisu and G.~Rizzoni, ``A comparative study of supervisory control strategies
  for hybrid electric vehicles,'' {\em IEEE transactions on control systems
  technology}, vol.~15, no.~3, pp.~506--518, 2007.

\bibitem{bertsekas1995dynamic}
D.~P. Bertsekas, {\em Dynamic programming and optimal control}, vol.~1.
\newblock Athena scientific Belmont, MA, 1995.

\bibitem{bauer2018distance}
L.~P. Bauer, ``Distance-based optimization of 48v mild-hybrid electric
  vehicle,'' Master's thesis, The Ohio State University, 2018.

\end{thebibliography}

%

%
%
%




\end{document}